\documentclass{aa}
\usepackage{gensymb}
\usepackage{enumitem}

\usepackage{textcomp}
\usepackage{multirow}
\usepackage{xcolor}
\usepackage{txfonts}
\graphicspath{{img/},{../img/}}
\usepackage{xspace}
\usepackage{natbib}
\usepackage{graphicx}

\DeclareMathAlphabet{\mathcal}{OMS}{cmsy}{m}{n}
\SetMathAlphabet{\mathcal}{bold}{OMS}{cmsy}{b}{n}

\newcommand{\OC}[1]{\mathcal{O}\left(#1\right)}

\newcommand{\eq}[1]{Eq.\,(\ref{#1})}

\newcommand{\fig}[1]{Fig.\,\ref{#1}}

\newcommand{\sect}[1]{Sect.\,\ref{#1}}

\def\taurex{TauREx\xspace}
\def\pytmo{Pytmosph3R\xspace}
\def\exok{\texttt{Exo\_k}\xspace}
\def\wasp{WASP-121b\xspace}

\def\Dl{\Delta\ell}

\def\rc{r}

\def\rs{\rho}

\def\Lon{\lambda}
\def\lon{\Lon_0}
\def\Lat{\varphi}
\def\lat{\Lat_0}
\def\lonobs{\lambda_\mathrm{0,obs}}
\def\latobs{\varphi_\mathrm{0,obs}}

\def\lonstar{\lambda_\star}
\def\latstar{\varphi_\star}

\def\nlayers{N_{z,\mathrm{lay}}}
\def\nplayers{N_{p,\mathrm{lay}}}
\def\nlevels{N_{\mathrm{lev}}}
\def\nslices{N_{\alpha}}
\def\nlat{N_{\mathrm{lat}}}
\def\nlon{N_{\mathrm{lon}}}
\def\nradial{N_{r}}
\def\ntheta{N_{\theta}}

\def\nwl{N_{\lambda}}
\def\nmol{N_\mathrm{mol}}

\def\tr{\tau_{\rc,\theta}^\lambda}

\def\trans{\mathcal{T}_{\rc,\theta}}
\def\tr{\tau_{\rc}^\lambda}

\def\trans{\mathcal{T}_{\rc}}

\newcommand*\diff{\mathop{}\!\mathrm{d}}

\newcommand{\balign}[1]{
\begin{align}
#1
\end{align}}

\usepackage[caption=false]{subfig}

\usepackage{float}
\floatstyle{plaintop} 
\restylefloat{table}

\usepackage{pgfplotstable}
\pgfplotsset{compat=1.14}
\pgfplotstableset{
 col sep=comma,
  empty cells with={\ensuremath{-}},
 empty cells with={},
 every head row/.style={before row={\hline},after row=\hline,},
 every last row/.style={after row=\hline},
}
\pgfplotstableset{
 multistyles/.style 2 args={
    @multistyles/.style={display columns/##1/.append style={#2}},
    @multistyles/.list={#1}
 }
}
\usepackage{hyperref}
\hypersetup{colorlinks=true,citecolor=blue,linkcolor=blue!80!black,urlcolor=blue!80!black}

\begin{document}

\title{Toward a multidimensional analysis of transmission spectroscopy.
Part I: Computation of transmission spectra using a 1D, 2D, or 3D atmosphere structure}
\titlerunning{Computation of transmission spectra using a 1D, 2D, or 3D atmosphere structure}

\author{Aur\'{e}lien Falco\inst{1}
\and
Tiziano Zingales\inst{1, 2}
\and
William Pluriel\inst{1}
\and
J\'{e}r\'{e}my Leconte\inst{1}
}

\institute{Laboratoire d'Astrophysique de Bordeaux, Univ. Bordeaux, CNRS, B18N, all\'{e}e Geoffroy Saint-Hilaire, 33615 Pessac, France \and Universit\`a di Padova, Dipartimento di Astronomia, vicolo dell’Osservatorio 3, 35122 Padova, Italy}

\date{\today}

\abstract{Considering the relatively high precision that will be reached by future observatories, it has recently become clear that one dimensional (1D) atmospheric models, in which  the atmospheric temperature and composition of a planet are considered to vary only in the vertical, will be unable to represent exoplanetary transmission spectra with a sufficient accuracy.
This is particularly true for warm to (ultra-) hot exoplanets because the atmosphere is unable to redistribute all the energy deposited on the dayside, creating a strong thermal and often compositional dichotomy on the planet. This situation is exacerbated by transmission spectroscopy, which probes the terminator region. This is the most heterogeneous region of the atmosphere. However, if being able to compute transmission spectra from 3D atmospheric structures (from a global climate model, e.g.) is necessary to predict realistic observables, it is too computationally expensive to be used in a data inversion framework. For this reason, there is a need for a medium-complexity 2D approach that captures the most salient features of the 3D model in a sufficiently fast implementation. With this in mind, we present a new open-source documented version of \pytmo that handles the computation of transmission spectra for atmospheres with up to three spatial dimensions and can account for time variability. Taking the example of an ultrahot Jupiter, we illustrate how the changing orientation of the planet during the transit can allow us to probe the horizontal variations in the atmosphere.
We further implement our algorithm in \taurex to allow the community to perform 2D retrievals.
We describe our extensive cross-validation benchmarks and discuss the accuracy and numerical performance of each model.
}

\maketitle

\section{Introduction}
\label{introduction}

Transmission spectroscopy is a powerful method for studying exoplanetary atmospheres.
During the past decade, space missions and ground-based surveys have provided numerous data on exoplanetary atmospheres  \citep{Tinetti2007,Tsiaras2018,Seidel2019,Mikal-Evans2020}.

In addition, new space telescopes will be launched in the current decade (James Webb Space Telescope (JWST) and Ariel) whose goals will not only be the detection of molecular features, but also the estimation, with greater accuracy, of the atmospheric molecular abundances.
However, a higher precision in the observation must be followed by an improvement of theoretical and statistical tools.

Several global climate models (GCM) are available \citep{Showman2002,Showman2008,Menou2009,Wordsworth2011,LFC13b,Charnay_2015,Kataria2016,Drummond2016,TK19} that can describe very complex 3D atmospheric structures and might in principle give us deep insight into the physics of a planetary atmosphere.
GCM simulations are a useful tool for characterizing from a theoretical point of view the chemical composition of an atmosphere and its physical properties \citep{Showman2008,Leconte2013,Guerlet2014,Venot2014,parmentier2018}.
These simulations can also help validate the parameters inferred by retrieval procedures \citep{Irwin2008,Al-Refaie2019,molliere2019}.
Finally, it is also possible to use the 3D structure of these simulations to compute transmission spectra, as discussed in \citet{caldas2019effects}.

However, the computational cost of a 3D GCM simulation is very high.
Their usage as forward models in Markov chain Monte Carlo (MCMC) retrieval procedures is therefore not currently considered.
If we were to simplify the parameterization of the 3D structure of the planet (to avoid the cost of the full GCM simulation), then the question of the parameters and their number would also affect the computational cost of the retrieval.
Moreover, retrieving too many parameters might be an issue: it would create many degeneracies in which the information would be lost, which would make the results difficult to interpret.

Bayesian retrieval codes include a critical trade-off between the precision of a model and its computational cost \citep{Al-Refaie2019,Waldmann2015a,Waldmann2015b,Line2013,Irwin2008}.
To increase the reliability of the solution given with a Bayesian approach, it is crucial to be able to quickly generate a forward model, without sacrificing meaningful physical phenomena that could lead to a significant spectral contribution.
So far, models that were used to infer atmospheric parameters in transmission or emission spectroscopy within Bayesian frameworks have mainly relied on a one-dimensional structure.
The use of such models was partly justified by the low precision in the available observational data \citep{Stevenson2014,Line2016,Tsiaras2018,Edwards-ares2020,pluriel-ares2020}.
In some cases, 1D models are a good approximation because the features in the transmission spectra come from a thin annulus around the limb, so that the region probed is almost homogeneous \citep{Barstow2017,Tsiaras2018,Guilluy2021,Swain2021}.

However, 1D atmospheric models will hardly explain the spectral shapes detected with the new generation of instruments because they reveal physical and chemical effects due to the 3D geometry of the atmosphere \citep{caldas2019effects,Changeat_2019,MacDonald2020,pluriel2020strong}.
In particular, \citet{pluriel2020strong} showed that the atmospheric parameters of ultra hot Jupiters retrieved using 1D models can be biased: the solution that fits the observation best could be very different from the reality.
In the case of ultra hot Jupiters, 1D retrieval codes cannot fit transmission spectra well because of the significant day to night thermal and chemical dichotomy.
In these atmospheres, the 1D vertical assumption is no longer valid because the region that is probed extends significantly across the limb on both the day- and the nightside of the planet.
The transmission spectra carry the information of the absence of water on the dayside (and its presence in the colder nightside) due to the thermal dissociation, and
the presence of strong CO features (which does not dissociate because of its stronger triple bond \citep{Lodders2002}).
A 1D retrieval will try to fit the water on the nightside (where it is not dissociated) and retrieves a colder temperature.
It compensates for this low temperature by overestimating CO to try to fit the CO features.

From this observation, the effect of the large difference between a hot temperature on the dayside and a cold temperature on the nightside may be expressed with a simpler model that is composed of only two dimensions: (1) a vertical dimension, and (2) an angular dimension following the star-observer axis.
Nevertheless, there are two caveats to keep in mind for the transition from 1D to 2D retrievals.
First, the computational time needed to converge using MCMC or nested sampling would be increased.
Second, and more importantly, we need to be aware that increasing the number of parameters in these retrieval methods could bring more degeneracy than information because the parameter space that is to be explored becomes larger.
Therefore we need a 2D geometry with the simplest parameters space, so that we can reproduce the 2D effects of the atmosphere without adding too many parameters that are to be retrieved.

In this first paper of a series, we present our general method for computing transmission spectra using atmospheric simulations with a different number of spatial dimensions and validate the numerical tools that we have developed for this purpose. We first introduce
a new open-source documented version of \pytmo \citep{caldas2019effects} that computes transmission spectra for atmospheres with up to three spatial dimensions. The code is very flexible and can use 3D time-varying atmospheric structures from a GCM as well as simpler, parameterized 1D or 2D structures. \sect{sec:application} shows some application examples, highlighting the benefits of realistically describing the complex structure of the atmosphere.
To allow our approach to be used in a retrieval framework, we have implemented our 2D algorithm in \taurex \citep{Waldmann2015a,Al-Refaie2019,Changeat_2019}. We then demonstrate that our implementation is sufficiently fast to allow converging on a realistic retrieval solution in a reasonable amount of time.

\section{Model description}
\label{model_description}

\begin{figure*}\centering
        \subfloat[\wasp (GCM)]{%
                \includegraphics[width=.24\textwidth,height=3cm,keepaspectratio]{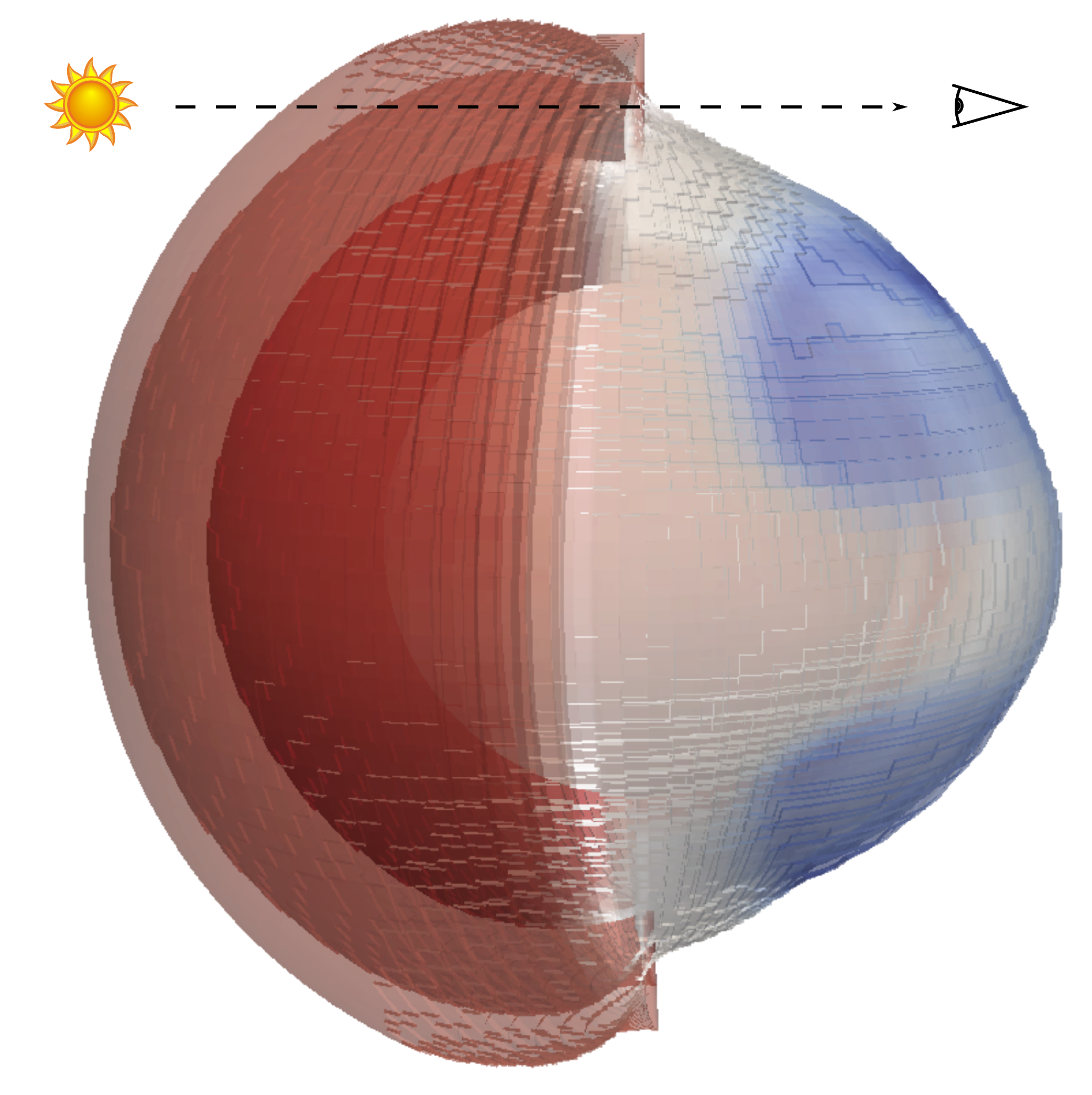}
                \label{WASP}
}\qquad
        \subfloat[3D]{%
                \includegraphics[width=.24\textwidth,height=3cm,keepaspectratio]{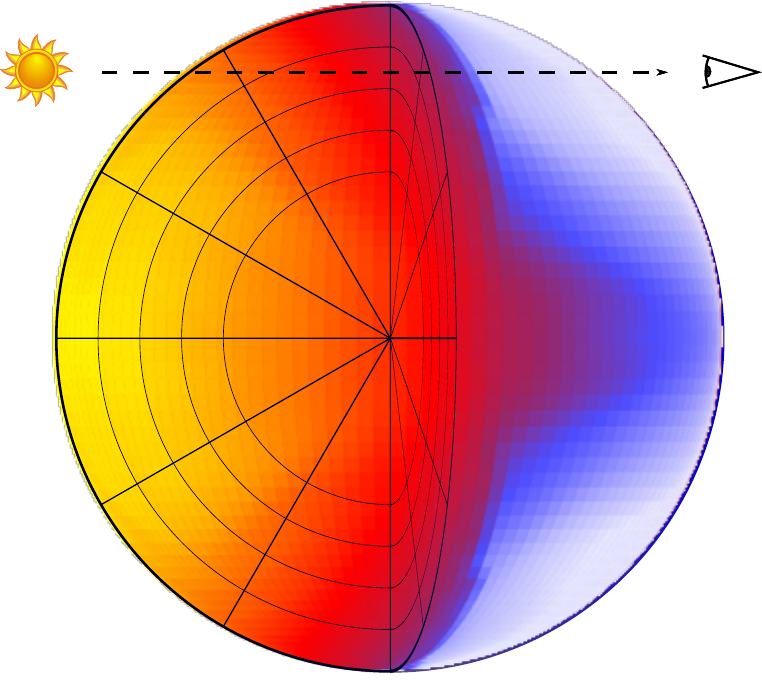}
                \label{3D}
}\qquad
        \subfloat[2D]{%
                \includegraphics[width=.24\textwidth,height=3cm,keepaspectratio]{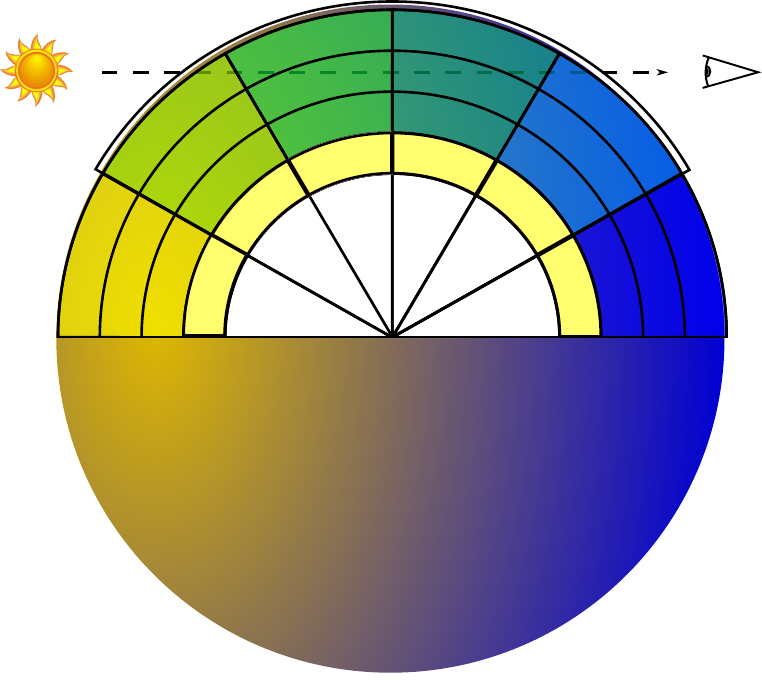}
                \label{2D}
}\qquad
    \subfloat[1D]{%
                \includegraphics[width=.24\textwidth,height=3cm,keepaspectratio]{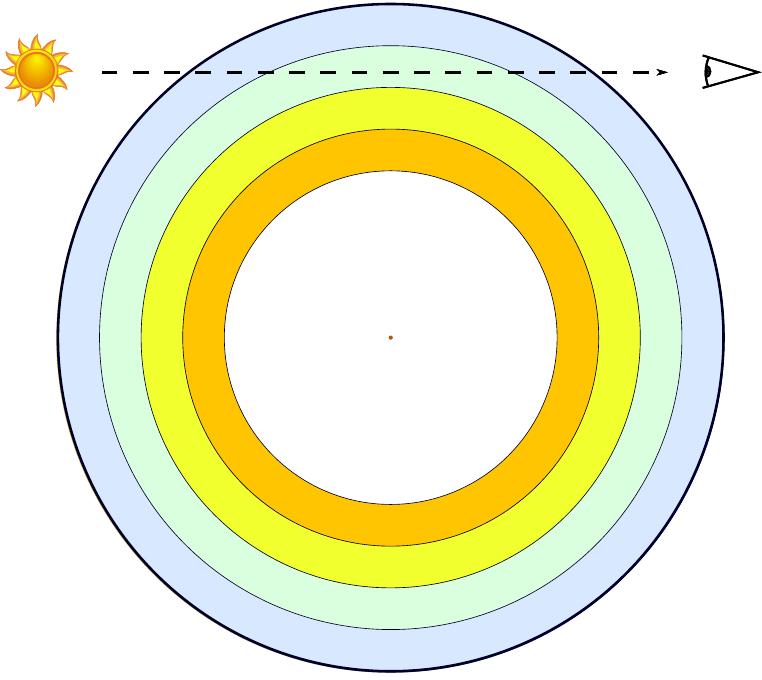}
                \label{1D}
}\qquad
        \caption{Visual representation of the dimensions of the models we considered
        using an equatorial view seen from the east.
        The colors of \protect\subref{WASP} indicate the temperature (redder is hotter) at isobar levels.
        The higher temperature on the dayside (on the left) affects the scale height, enlarging the atmosphere.
        The 3D model \protect\subref{3D} follows a spherical (radius, latitude, longitude) coordinate system, while the 2D model \protect\subref{2D} uses a polar grid whose the radial axis is the altitude and whose angular axis follows the solar altitude angle (which incidentally is also the angle of the zenith of the considered point with the zenith of the terminator). The 1D model \protect\subref{1D} simply relies on the altitude.
        }
        \label{models_dimensions}
\end{figure*}

We discuss here the computation of transmission spectra in the case of 3D, 2D, or 1D simulations.
A simple representation of each model is shown in \fig{models_dimensions}.
An example of a 3D GCM simulation, representing \wasp, is included. It shows that the higher temperature on the dayside (on the left) affects the scale height, enlarging the atmosphere.
We validate and study the performance of each configuration in Sec. \ref{model-validation} and \ref{model-time}.

\subsection{Three-dimensional case}
\label{3D_case}

The computation of the optical depth, and eventually the transmission spectrum in the case of 3D simulations, has already been discussed in \citet{caldas2019effects}.
We reintroduce the method and notations needed hereafter in the context of a new version of the implementation,
\href{https://forge.oasu.u-bordeaux.fr/jleconte/pytmosph3r-public}{\pytmo~2.0}, which is more robust and user friendly, of which the documentation is available \href{http://perso.astrophy.u-bordeaux.fr/~jleconte/pytmosph3r-doc/index.html}{here}
\footnote{\url{http://perso.astrophy.u-bordeaux.fr/~jleconte/pytmosph3r-doc/index.html}
\label{pytmodoc}
}.

Global climate model simulations such as the LMDZ GCM \citep{Wordsworth2011} provide a description of physical properties such as pressure, temperature, and volume-mixing ratios of absorbing or scattering molecules and aerosols in a three-dimensional grid.
The vertical dimension of this grid may rely on pressure levels, for example, while the horizontal grid relies on latitude and longitude coordinates.
The number of pressure levels is noted $\nplayers$, the number of latitudes $\nlat$ , and the number of longitudes $\nlon$.

\pytmo~2.0 offers several options to compute the volume-mixing ratio of all gases in each cell of the atmospheric grid: (1)~It can extract this information directly from the input simulation when it is present, (2)~it can interpolate from a separate table providing the mixing ratios on a pressure and temperature grid (e.g., when thermodynamical equilibrium is assumed), (3)~it can use analytical formulae (e.g., those of \cite{parmentier2018} where thermal dissociation of key molecules is accounted for), (4)~it can call a chemistry module like FastChem \citep{stock2018fastchem} to compute the abundances on the fly, and (5)~the users can specify their own constant mixing ratios.
Other types of chemistry may be easily implemented by the users to adjust to the context of the simulation.

As discussed in \citet{caldas2019effects}, because i) isobaric surfaces are generally not iso-altitude surfaces and ii) altitude is the key variable in transmission geometry, the input simulation needs to be interpolated into a
(radius, latitude, longitude)
spherical coordinate system. The vertical axis of this new grid is discretized in $\nlayers$ layers
delimited
by $\nlayers+1 \equiv\nlevels$ points called levels.
The number of latitudes and longitudes of the new grid is identical to that of the input grid.
The coordinates of the points in this new system are noted $(\rs,\lat,\lon)$.

The user can specify the position of the observer in this latitude-longitude system $(\latobs,\lonobs)$. This can be used to study the spectral variations caused by the rotation of the planet during the transit, as discussed in Sec.~\ref{sec:application}.

\label{rays_definition}
\label{position_observer}
The optical depth is computed for a set of rays that are parallel to the planet-observer axis. As described extensively in \citet{caldas2019effects}, each ray is uniquely defined by its intersection with the plane normal to the planet-observer axis and passing through the planet center (hereafter, the plane of the sky). In this plane, the rays are
arranged on a polar grid with $\nradial$ points along the impact parameter axis that are uniformly distributed between the surface of the planet and the top of the atmosphere, and $\ntheta$ points along the azimuthal coordinate that goes around the limb of the planet.
Even though the number of radial points $\nradial$ in this grid of rays may be different from the number of $\nlayers$ by definition, they are ultimately connected.
Adding too many rays per layers does not increase the precision of the model because the information contained in the simulation does not increase, while too few rays would mean to lose some of this information.
We use $\nradial = \nlayers$ in this paper and discuss the position of the rays in the layers in Sec.~\ref{tests1D}.

This new version relies on \exok \citep{leconte2020spectral} for the interpolation of the opacities (including regular molecular and atomic transitions, collision-induced absorptions, Rayleigh scattering by the gas, and Mie scattering by aerosols). As a result, the user can use both high-resolution cross sections or correlated-$k$ coefficient tables.

\subsection{Two-dimensional case}
\label{2D_case}

To reduce the cost of the computation of a transmission spectrum, we may simplify the simulation by assuming symmetry around the star-observer axis.
We may thus reduce the simulation over a 2D grid that
includes
this axis.
This allows us to describe the day to night temperature differences of certain types of planets such as ultra hot Jupiters with a lower cost than a full 3D simulation, and with a better precision than 1D models \citep{pluriel2020strong}.
Because this model is entirely new (although it is based on the same method as the 3D model), we describe the details of its computation here.

\subsubsection{Input data}
\label{input_simulation}

The two-dimensional grid that we consider in this paper is
centered on the planet center and contains the star-observer axis.
Because atmospheric flows tend to follow isobars, the vertical dimension follows pressure levels, as is usually done in GCM simulations.
The angular dimension follows the solar altitude angle, and the star is placed
in the equatorial plane.

This 2D model could be applied on a complex 2D temperature, but we rely on the following thermal structure in this paper.
The temperature is defined for every level and angle $(i,\alpha^*)$ using
\setlength\arraycolsep{1pt}

\begin{equation}
\label{temperature2D}
\left\{ \begin{array}{lll}
        P > P_{iso}, & T = & T_{deep},\\
        \multirow{3}{*}{$ 
        P < P_{iso},
        \left\{ \begin{array}{l}
                2\alpha^* \geq \beta , \\
                2\alpha^* \leq -\beta, \\
                -\beta < 2\alpha^* < \beta,
        \end{array} \right. %
        $} 

        & T = & T_{day},\\
        & T = & T_{night},\\
        & T = & T_{night} \\
        & & + (T_{day} - T_{night}) 
        \frac{\alpha^*+\beta/2}{\beta},
\end{array} \right.
\end{equation}

\noindent where $T_{day}$, $T_{night}$ are scalar parameters that define temperatures of the day- and nightside.
The $\beta$ angle defines the area around the terminator, where the temperature decreases linearly from $T_{day}$ to $T_{night}$.
$T_{deep}$ defines the isothermal structure of the atmosphere for pressures higher than $P_{iso}$.
This simple parameterization of the temperature structure is inspired by GCM simulations of hot and ultra hot Jupiters \citep{Showman2015,parmentier2018,TK19} where the models can be approximated according to \eq{temperature2D}.
More particularly, \citet{pluriel2020strong} have shown that this 2D representation could approximate ultra hot Jupiters with a satisfying precision.

Like in the 3D model (see \sect{3D_case}), the chemical composition of the gas can parameterized by the user or computed using more complex chemical models.

\subsubsection{Interpolation into an altitude-based polar grid}
\label{altitude_based_grid}

\begin{figure}[ht]\centering
        \subfloat[Input grid. The pressure levels separating the $\nplayers$ layers are identical in each column. Because of the temperature and composition differences, the altitude of a layer changes from one column to the next.
        ]{%
        \includegraphics[width=.45\textwidth,keepaspectratio]{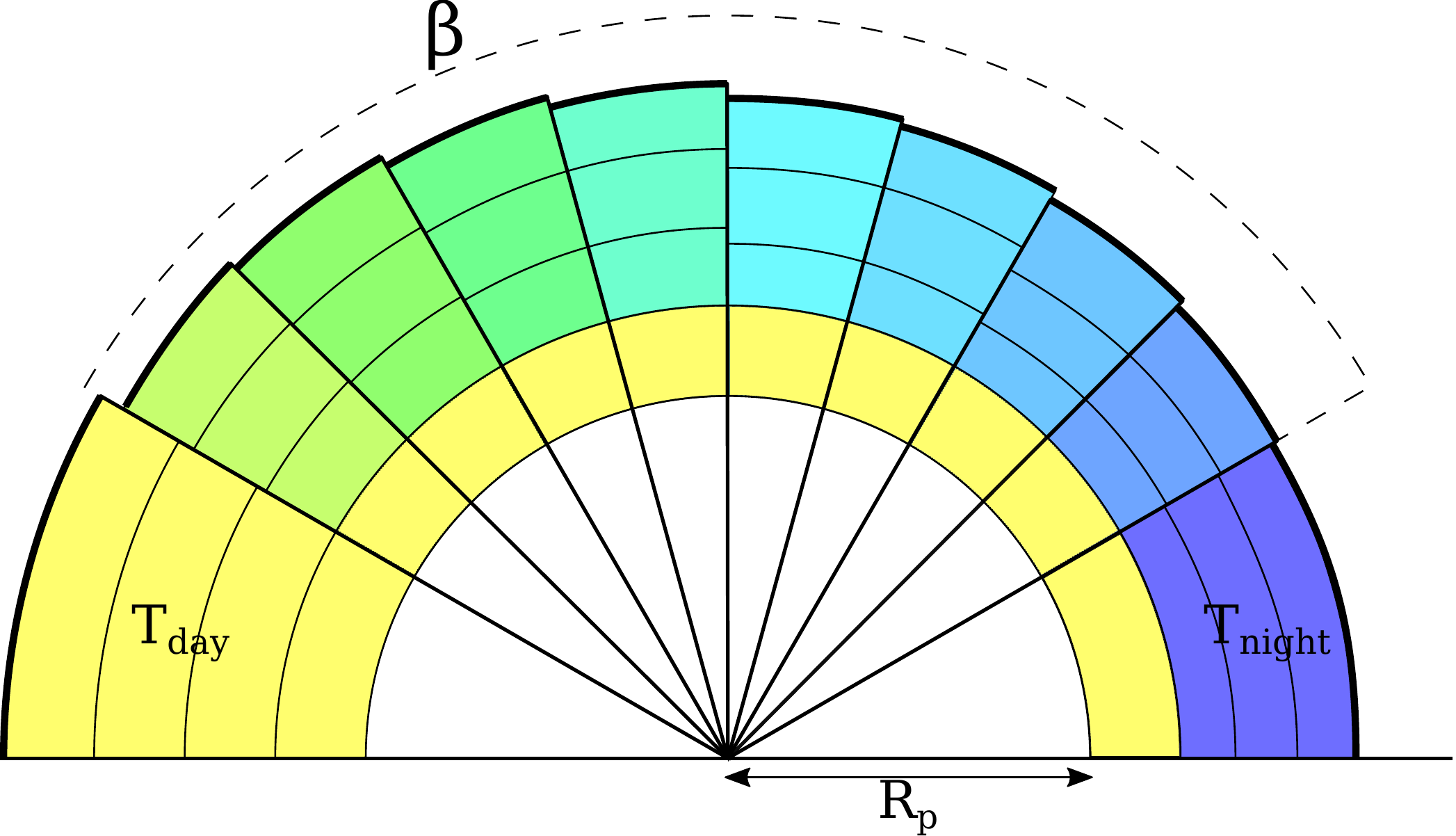}%
                \label{input_grid}
}\qquad
        \subfloat[Altitude-based 2D grid (in red) with $\nlayers$ layers superimposed on the input grid. In this example, we have $\nlayers=\nplayers$, but this is not a requirement.]{%
        \includegraphics[width=.45\textwidth,keepaspectratio]{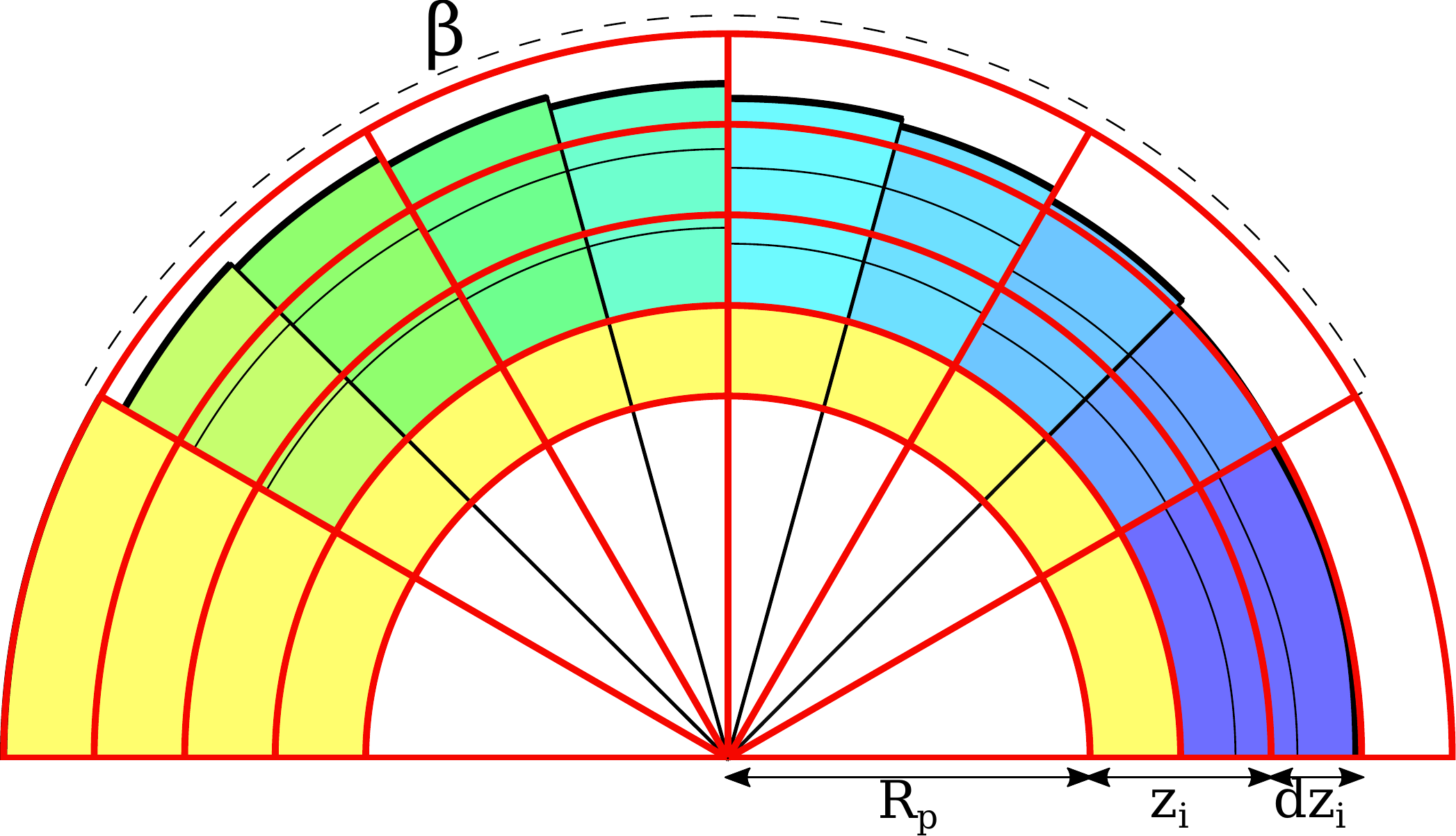}%
                \label{altitude_grid}
                }\qquad
        \caption{Illustration of the two grids used in the model. On the left, the pressure-based input grid \protect\subref{input_grid}. On the right, the regular altitude grid \protect\subref{altitude_grid}. The location of each of the $\nslices$ columns is parameterized by its solar elevation angle, $\alpha^*$, equal to 90$^\circ$ at the substellar point, 0$^\circ$ at the terminator, and -90$^\circ$ at the subobserver point. $\beta$ is the angle over which the atmosphere transitions from dayside to nightside temperatures.
        $R_p$ is the planetary radius at the bottom of the grid.
        $z_i$ is the altitude of level $i$, and $dz_i$ is the thickness of layer $i$ between levels $i$ and $i+1$.
        }
        \label{altitude_interpolation}
\end{figure}

To compute the intersections of the rays with the atmospheric grid (discussed in \sect{rays_intersections}), a regular geometric grid is to be preferred.
Because every grid point in our 2D
input grid can have different physical and chemical properties (pressures, temperatures, abundances, etc.) and hence different scale-heights, however, the altitude of the $n$-th layer will change from one column (parameterized by the angle $\alpha^*$) to the next, as illustrated by
\fig{altitude_interpolation}.

The altitude can be calculated from the hydrostatic equilibrium through the hypsometric equation,
\begin{gather}
        dp = -\rho \cdot g \cdot dz,\\
        z_{i+1} = z_{i} + \frac{M\cdot R\cdot T_{i}}{g(z_{i})} \cdot \ln\left(\frac{P_{i}}{P_{i+1}}\right),
        \label{hydrostatic}
\end{gather}
where $\rho$ is the mass density, $M$ is the molar mass of the gas, $R$ is the universal gas constant, and $g(z_{i})$ is the gravity at altitude $z_{i}$.
From this equation, we can infer that a difference of temperature in the angular dimension will indeed affect the scale height.

With the altitude computed in the input grid, we can construct a new grid based on the altitude, as is represented in \fig{altitude_grid}, in a linear discretization of $\nlevels$ points that define $\nlayers = \nlevels-1$ layers, up to a maximum altitude of our choice.
The data in this new polar grid can be interpolated using a logarithmic interpolation for the pressure and a linear interpolation for the temperature and chemistry.
This new grid facilitates computing the intersection points between the rays and the grid, as we discuss in Sec.~\ref{rays_intersections}.

The number of slices or angles in the 2D model is $\nslices$, of which $\nslices-2$ discretize the angle between $-\beta/2$ and $\beta/2$.
The first and last angular slices account for the day- and nightside (not subdivided as they are uniform).
The points in this 2D altitude-based coordinate system are noted hereafter $(\rs, \alpha)$.

\subsubsection{Intersections of the rays with the altitude grid}
\label{rays_intersections}

\begin{figure}[ht]\centering
        \includegraphics[width=.475\textwidth,keepaspectratio]{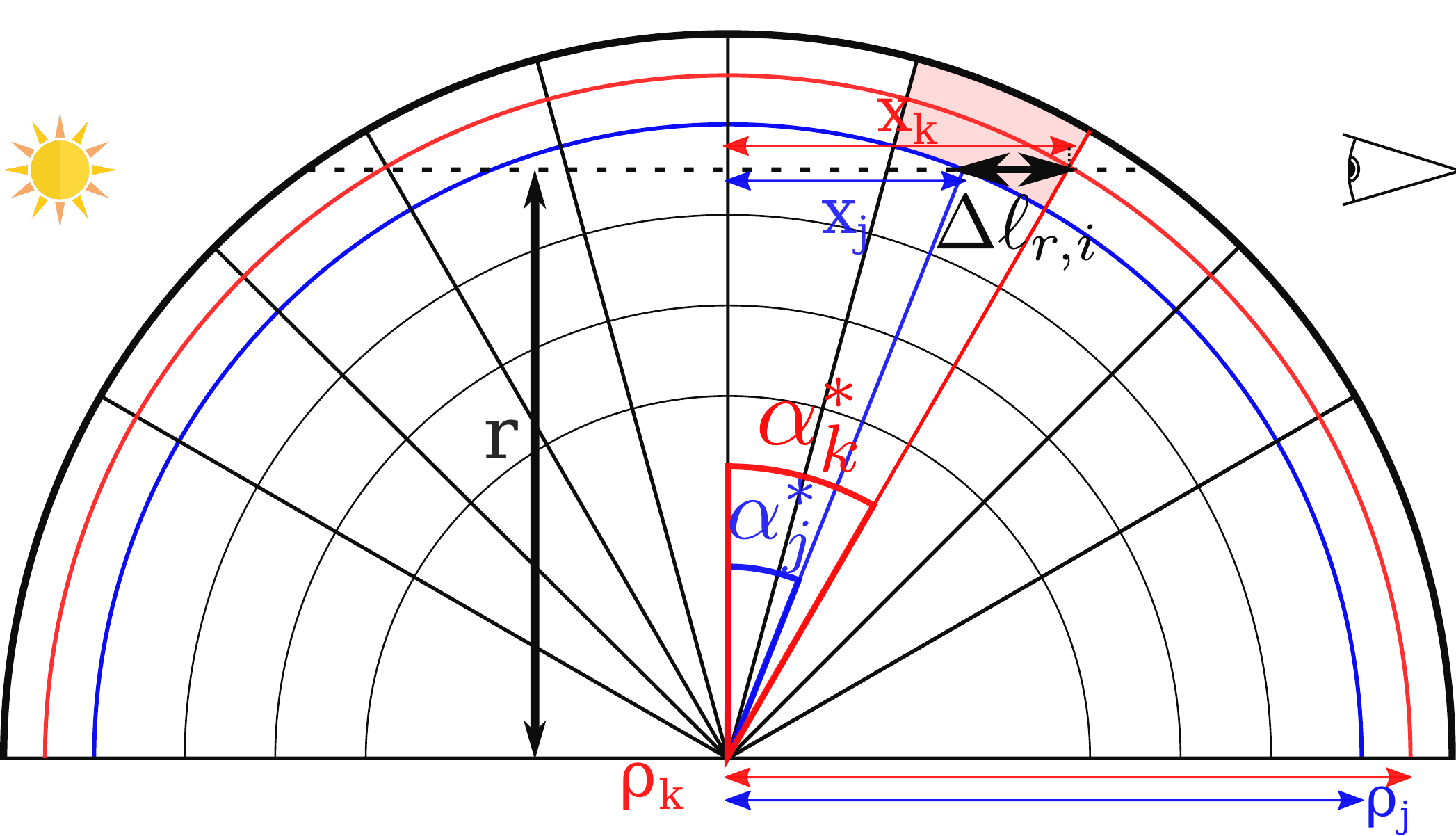}
        \caption{Intersection of a ray with the 2D grid.
        An example of intersections for one level and one angle is highlighted by points $j$ and $k$ at coordinates $(\rho_j, \alpha^*_j)$ and $(\rho_k, \alpha^*_k)$, respectively.
        The length $\Dl_{\rc,i}$ of the corresponding segment is deduced from the computation of the distances $x_j$ and $x_k$ of both points to the terminator.
        The coordinates of the segment must also be computed to obtain the physical properties of the cell.
}
\label{paths}
\end{figure}

In this model, we consider $\nradial = \nlayers$ rays.
A ray may be described using its impact parameter $\rc$, defined as the normal distance between the ray and the center of the planet (see \fig{paths}). We chose the grid of impact parameters so that the rays cross the plane of the sky (or equivalently here, the terminator of the planet) exactly at the center of the atmospheric layers.
To compute the optical depth (Eq. \ref{eq:tau}) of a ray $\rc$, we need the length of each segment of the ray crossing individual cells of the altitude grid, as well as the physical properties of these cells.

The coordinates of the intersections points of the ray with the grid may be computed using
\begin{equation}
        \rc = \rs \cos(\alpha^*).
        \label{eq:2D_intersection}
\end{equation}
This returns the angle of intersection when applied on the $\nlevels$ levels of radius $\rs$ (duplicated for positive and negative values), and the radius of intersection when applied on the $\nslices-1$ solar elevation angles $\alpha^*$ that separate the slices of the grid.
The two types of intersections (with levels and with angles) are visually represented in blue and red (or indices $j$ and $k$), respectively, in \fig{paths}.

The intersection points of a given ray are then sorted using their angular coordinate. The length of each segment (as highlighted in \fig{paths}) is equal to the subtraction of the distance of both extremities of the segment ($j$ and $k$ in our example) to the terminator.
The distance of a point $k$ to the terminator is given by
\begin{equation}
\mathbf{
        x_k^2 = \rs_k^2 - \rc^2.
        }
\label{eq:1D_intersection}
\end{equation}
This equation returns two solutions, corresponding to a point before and a point after the terminator (following the sign of the solution).
By convention, a negative distance corresponds to a point on the dayside.

We note $\Dl_{\rc,i}$ the distance of a segment $i$ between two consecutive intersection points of a ray $\rc$.

\subsubsection{Optical depth}
\label{optical_depth}

Based on the previous definitions, the optical depth of each ray $\rc$ at a wavelength $\lambda$ can be computed using Eq. \ref{eq:tau},

\begin{equation}
        \tr = \sum_{\substack{i}}
        \frac{P_{\rc,i}}{k_B T_{\rc,i}}
        \left(\sum_{m=1}^{N_{gas}}
        \chi_{m,\rc,i} \sigma_{m, \lambda} +
        \sum_{j=1}^{N_{con}} k_{mie, j}
        \right)\Dl_{\rc,i},
        \label{eq:tau}
\end{equation}
where $P_{\rc,i}$ and $T_{\rc,i}$ are the pressure and temperatures of the cell corresponding to the segment $i$ of the ray $\rc$, $k_B$ is the Boltzmann constant, $\chi_{m,\rc,i}$ is the volume-mixing ratio of the
$m$-th
molecule, $\sigma_{m, \lambda}$ is the total cross-section of Rayleigh scattering and molecular and continuum absorptions, and finally, $k_{mie, j}$ is
the extinction coefficient associated with the Mie scattering for the $j$-th aerosol.
$N_{gas}$ is the number of molecules, and
$N_{con}$ is the number of aerosols.
The cross-sections and Mie coefficients can be computed by \exok \citep{leconte2020spectral}.

\subsubsection{Transmittance map and spectrum}
\label{spectrum}

A transmittance map may be computed from the optical depth through
\begin{equation}
        \trans^\lambda = e^{-\tr} .
        \label{eq:transmittance}
\end{equation}
In the case of a homogeneous stellar disk, the relative dimming of the stellar flux is given by
\begin{equation}
        \Delta_\lambda = \frac{{\pi R_p}^2 + \sum\limits_{\rc} \left(1 - e^{-\tr}\right) S_\rc }{{\pi R_s}^2},
        \label{eq:integral}
 \end{equation}

with
$S_\rc = 2 \pi (\rc + \frac{\diff\rc}{2})\diff\rc$.
$R_p$ is the radius of the planet, $R_s$ is the radius of the star, and $\diff\rc$ is the distance between two consecutive $\rc$.
The computation of the transmission spectrum is complete.

\subsection{One-dimensional case}
\label{1D_case}

To simplify the simulation even further, we can approximate the model using only one dimension: the vertical axis. This approximation has been extensively used in the past and will serve as a reference.

In this case, the temperature and composition are horizontally uniform, so that isobaric and iso-altitude surfaces are equivalent and a single grid can be used.
The steps required to compute the transmission spectrum are then exactly identical to those in the 2D model, except that we do not need to use Eq.~\ref{eq:2D_intersection} because there is no slice in this model.

\begin{figure}[h]\centering
        \subfloat[Rays at levels]{%
        \includegraphics[width=.45\columnwidth,height=.6\textheight,keepaspectratio]{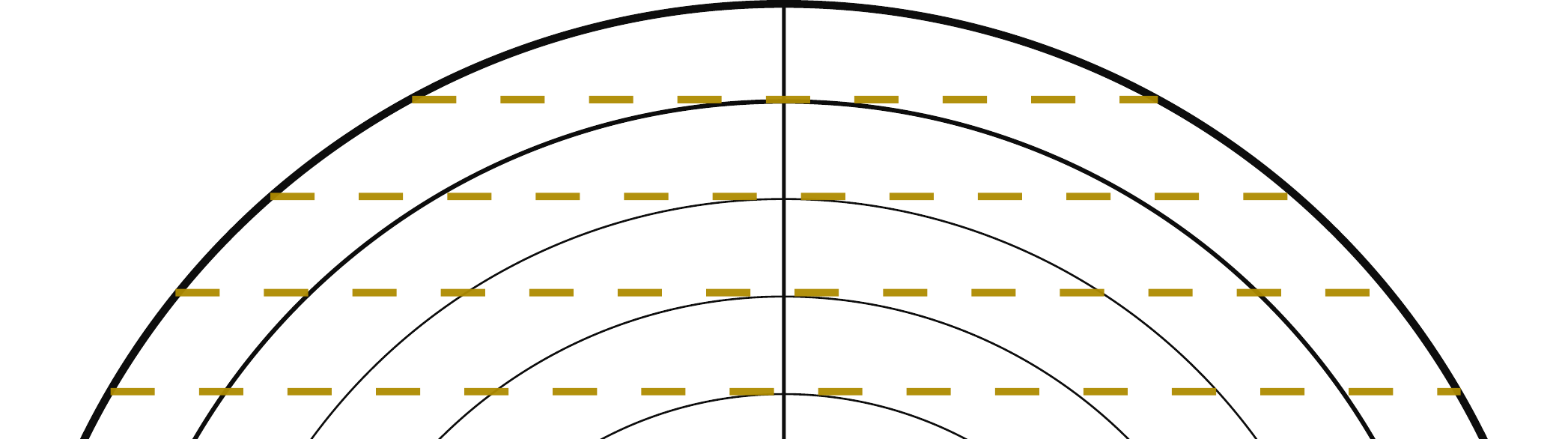}%
        \label{rays_bottom}
}\qquad
        \subfloat[Rays at mid-layers]{%
        \includegraphics[width=.45\columnwidth,height=.6\textheight,keepaspectratio]{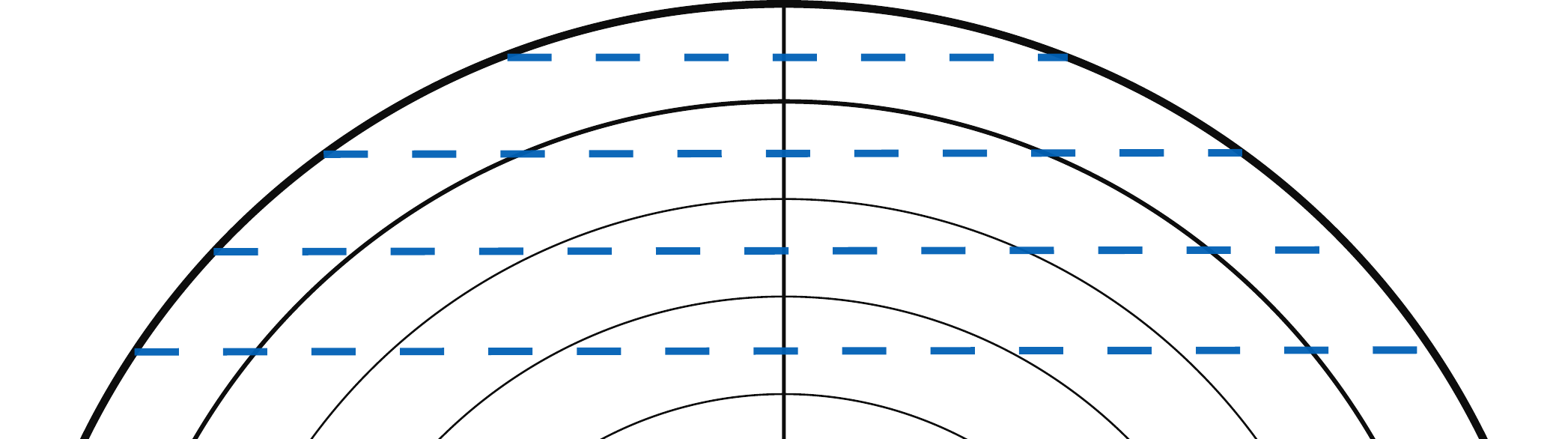}%
        \label{rays_center}
}\qquad
        \caption{Schematic of the two possible methods for positioning the rays (dashed lines). In panel \protect\subref{rays_bottom}, the rays are tangent to the pressure levels (solid arcs), as done in \taurex~3.0. In panel \protect\subref{rays_center}, the rays pass the middle of the layers, as done in \pytmo and throughout this paper. A comparison of the accuracy of these two methods is shown in \fig{iso_results}}
        \label{rays_position}
\end{figure}

However, although there is consensus about the theoretical equations for computing the transmission spectrum of a horizontally uniform atmosphere, some differences in the numerical algorithm can lead to significant quantitative differences in the results. In particular, there are two main options when the impact parameters of the rays are chosen, as shown in \fig{rays_position}.
The first option (used in \taurex~3.0) is to place the rays so that their impact parameter $\rc$ is equal to the radius of the levels $\rho$.
The second option is to place the rays so that their impact parameter $\rc$ is equal to the radius of the midpoints of the layers $\rho+\frac{\diff \rho}{2}$.

As we show in Sec.~\ref{tests1D}, the second option (rays at midlayers) appears to results in a faster convergence of the numerical scheme. We therefore implemented this method in our 2D and 3D codes.

\section{Model accuracy}
\label{model-validation}

In this section, we use a series of test cases of increasing complexity to cross-validate our different implementations.
As a byproduct, this will allow us to test the accuracy of our algorithms for various grid resolutions.
The efficiency in terms of computing time is discussed in \sect{model-time}.

We emphasize that our test cases are based on the atmosphere of an ultra hot Jupiter, \wasp, where hydrogen partially dissociates so that the scale height is extremely large.
In addition, the planet is extremely inflated. All these factors add up to increase the atmospheric signal up to several thousand ppms.
As a result, this is a particularly stringent test for models, which explains the relatively high resolution needed to achieve a given precision. A lower resolution could be used for cooler objects.

\subsection{Experimental setup}
\label{experimental_setup}

The 1D model is based on the \taurex 3 implementation \citep{Al-Refaie2019}.
We developed the 2D model in the same framework, extending each 1D data profile to a 2D alternative, and using Eq. \ref{temperature2D} for the temperature map.
The 3D model, developed in a first version by \citet{caldas2019effects} was reimplemented here in a second version,
\href{http://perso.astrophy.u-bordeaux.fr/~jleconte/pytmosph3r-doc/index.html}{\pytmo~2.0}$^{\ref{pytmodoc}}$.
All models are coded in Python 3, with computationally intensive operations relying on numba \citep{lam2015numba}.
The machines on which the numerical experiments were performed are Intel®  Xeon® Gold 6138 CPUs with 40 = 20$\times$2 cores @ 2.00GHz, with 128GB of RAM.

\subsection{Validation for the 1D test cases}
\label{tests1D}

A test case that can be reproduced by each model is the case of isothermal atmospheres.
We therefore study here an isothermal example of a hot Jupiter planet with the physical properties listed in Table~\ref{tab:isothermal}.
\begin{table}[h]\centering
    \begin{tabular}{|c|c|}
    \hline
        Planet radius &
        1.807 $R_J$
        \\
        Surface gravity &
        9.39 m.s$^{-2}$
        \\
        Temperature &
        2500 $K$ \\

        [H$_2$O] & $5.01 \cdot 10^{-4}$ \\

        [CO] &
        $4.4 \cdot 10^{-4}$        \\

        [H$_2$] &
        $0.740$        \\

        [He] &
        $0.259$
        \\\hline
    \end{tabular}
    \caption{Characteristics of the isothermal case, including the abundances (volume mixing ratios) of each molecule.
    }
    \label{tab:isothermal}
\end{table}
The spectrum is generated for 39124 wavelengths from 0.3 to 15~$\mu$m.

We have ensured that all codes indeed converge toward a solution when the number of layers is increased, as is shown by Fig. 4 of \citet{caldas2019effects} for the previous version of \pytmo.

\begin{figure}\centering
        \includegraphics[width=.5\textwidth]{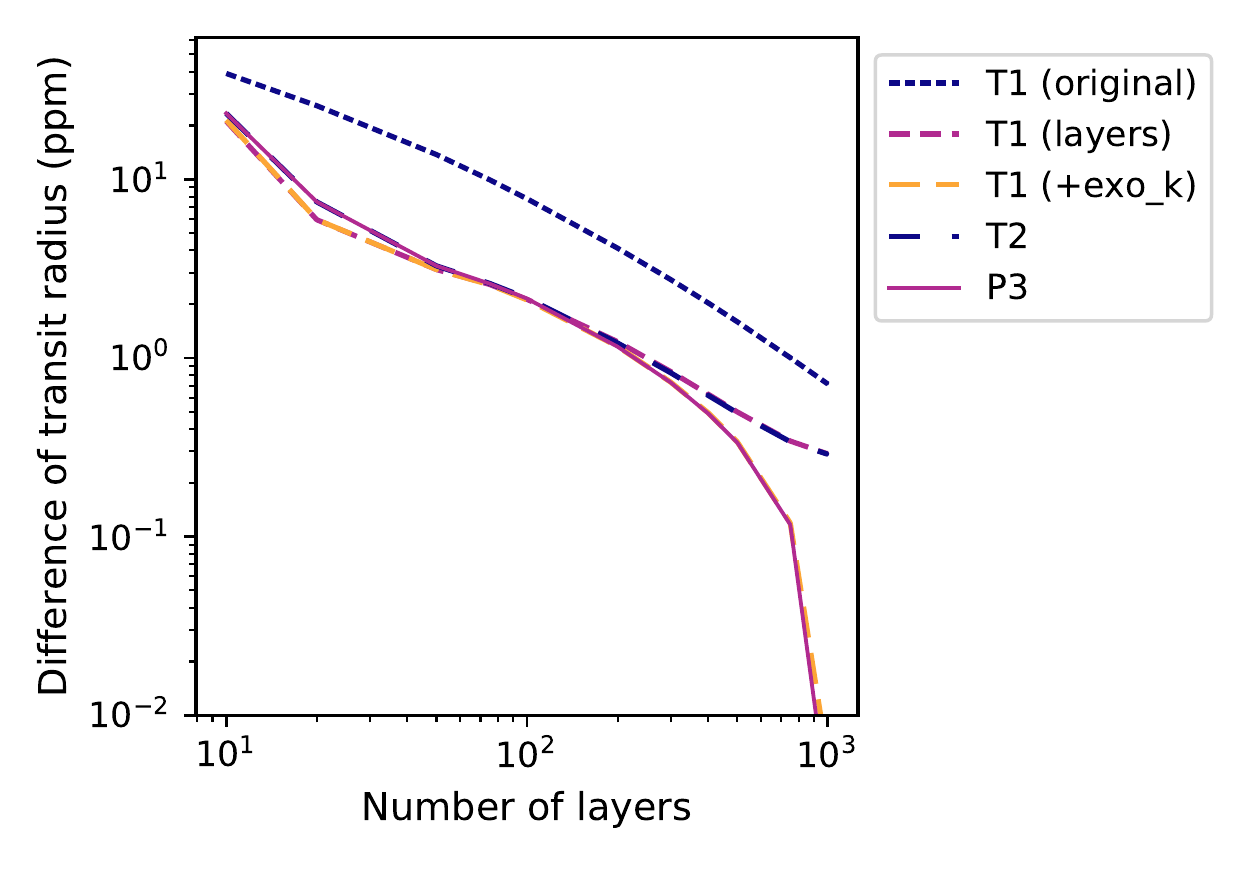}
        \caption{Average of the absolute difference (in ppm) between each model and the converged solution of \pytmo (with 1000 layers) as a function of the number of layers. P3 stands for \pytmo, and T1 and T2 for \taurex 1D and 2D, respectively. The original implementation of \taurex (denoted "T1 (original)", with rays that are tangent to levels) is included.
        }
        \label{iso_results}
\end{figure}

\fig{iso_results} shows the convergence of the \taurex and \pytmo 2.0 with respect to a converged \pytmo (1000 layers).

To accelerate the convergence of \taurex~1D, we implemented a new version that places the rays at the center of the layers (``T1 (layers)'') and not at the levels.
The original implementation of \taurex is given by the curve ``T1 (original)''.
See \fig{rays_position} for a visual representation of the two methods.
This simple algorithmic change improves the model accuracy by
a factor of 3 on average
at no cost.

For more than $\sim$1000 layers, the difference between T1 (layers) and our reference seems to reach a plateau.
When the interpolation of the opacities in \taurex\ is replaced by that of \exok (the one used in \pytmo) in addition to the position of the rays at the center of the layers, we obtain the method ``T1 (+exo\_k)'' for which the difference with the reference model continues to decrease as the number of layers increases.
This shows that in the case of relatively hot giant planets, the opacity interpolation scheme can lead to errors of $\sim0.3$ppm.

Overall, all (new) codes converge to a difference smaller than 1~ppm with a few hundred layers.
From this point forward, all methods rely on \exok for the computation of the opacities and place the rays at the center of the layers.

For horizontally uniform atmospheres, the choice of the orientation of the longitude or latitude grid used in \pytmo is arbitrary and should not affect the results. We verified that the output spectrum is independent of this choice down to machine precision, which further validates our implementation.

\subsection{2D test case}
\label{tests2D}

\begin{figure}\centering
                \includegraphics[width=0.45\textwidth]{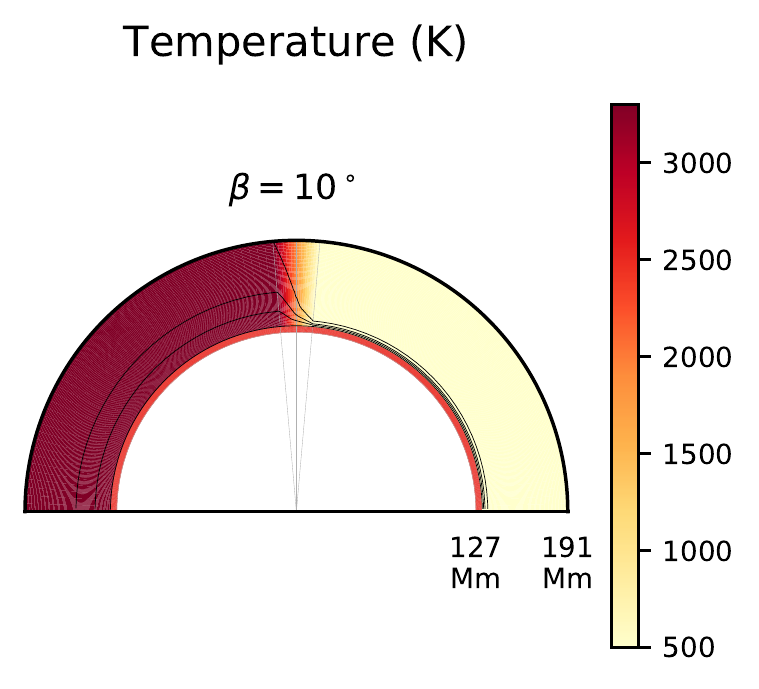}
                \caption{Temperature map for our reference 2D test case. The substellar point is on the left. $T_{day} = 3300K$ while $T_{night} = 500K$ and $T_{deep} = 2500K$, with $\beta = 10$\textdegree. The black lines correspond to isobar levels.
                The altitude is in Mm, i.e., thousands of km.
                }
                \label{temp2D}
\end{figure}

We now cross-validate \taurex~2D and \pytmo using a 2D temperature structure that is symmetrical around the planet-observer axis (see \eq{temperature2D}). The chosen temperature structure, shown in \fig{temp2D}, has a large difference between the dayside and nightside  temperature.

Our temperature structure does not \textup{\textit{\textup{directly}}} depend on the longitude and latitude: it only depends on the solar elevation angle, which is given by
\balign{
        \sin \alpha^* = &\sin \lat \sin \latstar\nonumber\\
        &+ \cos \lat \cos \latstar \cos(\lon - \lonstar),
        \label{solar_elevation_angle}
}
where $(\lat,\lon)$ are the latitude and longitude of the current cell, and $(\latstar,\lonstar)$ are the latitude and longitude of the substellar point.
As a result, the choice of the grid orientation is arbitrary. In other words, the choice of the direction of the star in our arbitrary reference frame is a free parameter that should not affect the results (as long as we align star, planet, and observer).

\begin{figure}\centering
        \subfloat[$\latstar$~=~0\textdegree]{%
        \includegraphics[width=.22\textwidth,height=.6\textheight,keepaspectratio]{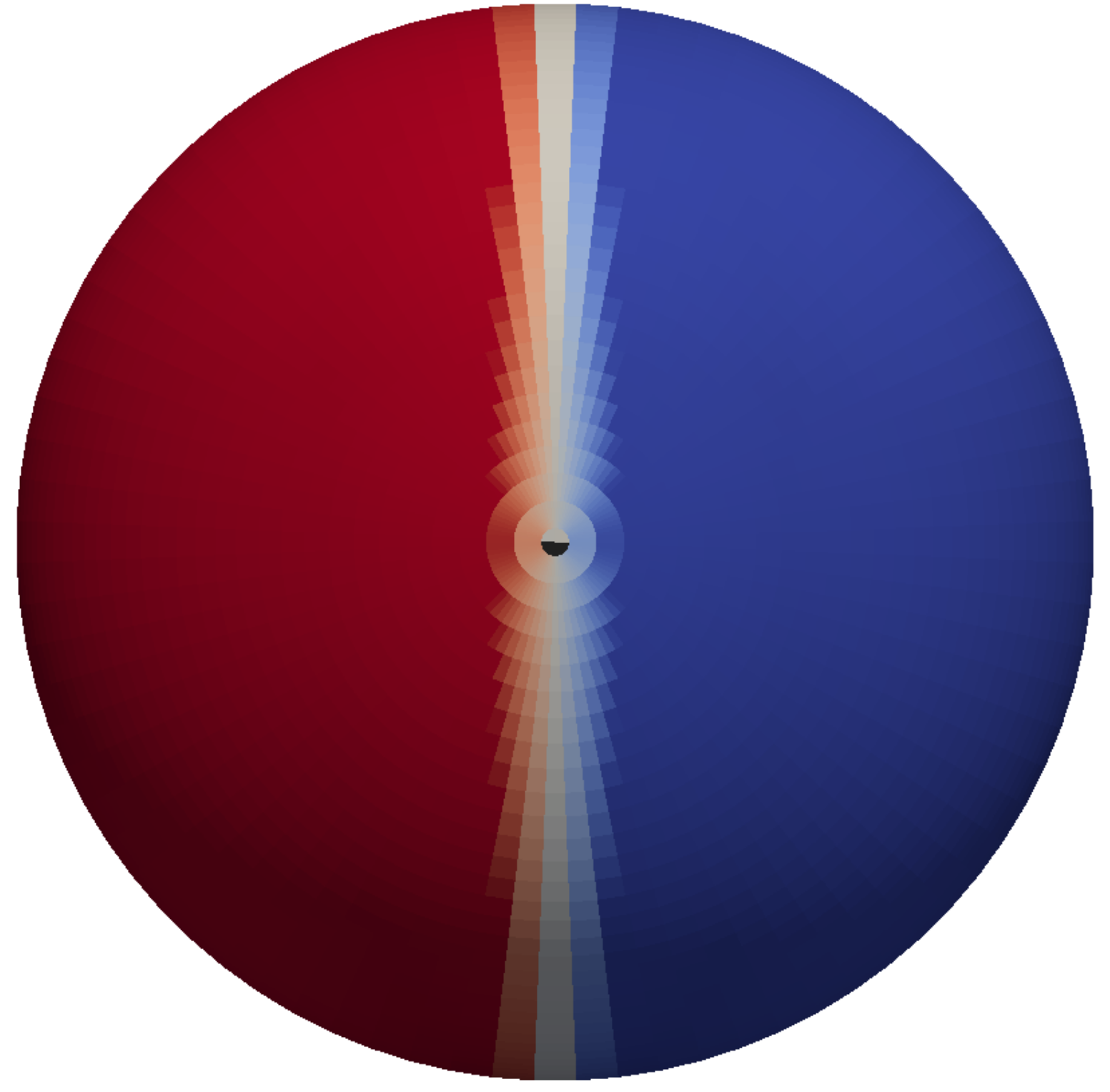}%
        \label{2d_equator}
}\qquad
        \subfloat[$\latstar$~=~90\textdegree]{%
        \includegraphics[width=.22\textwidth,height=.6\textheight,keepaspectratio]{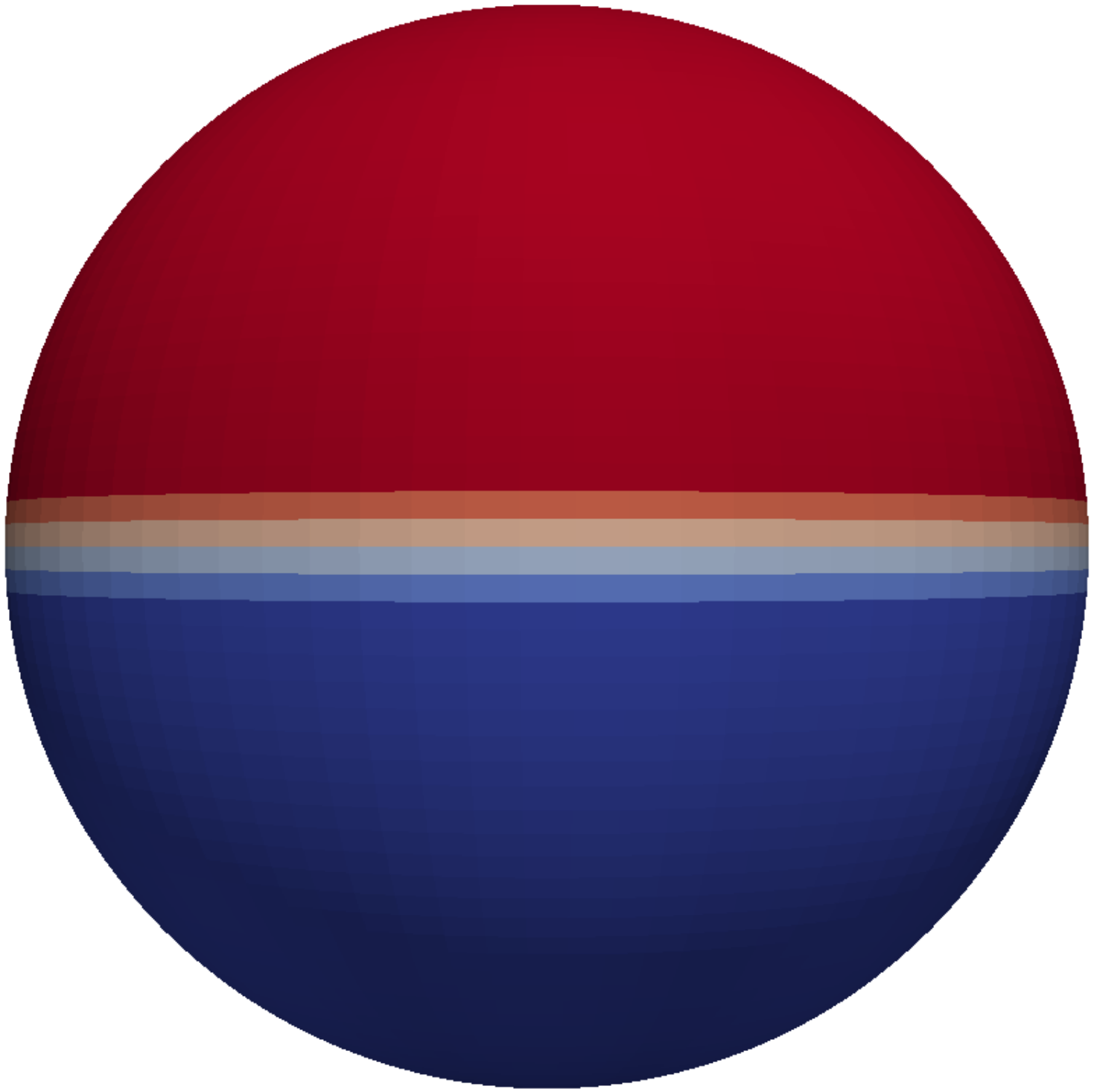}%
        \label{2d_pole}
}\qquad
        \caption{Discrete horizontal temperature maps
        at high altitude
        computed using \eq{temperature2D} with 80/80 latitude and longitude grid points with two different choices of grid orientation (redder is hotter).
        These maps are visualized in 3D through ParaView \citep{ahrens2005paraview}.
        a) The star is located
        in the equatorial plane (the planet is seen from the pole, with the star on the left).
        b) The star is located at the pole (the star and the north pole are located at the top).
        In this idealized setup, choice b allows us to take advantage of the symmetries of the system and to use only one longitude point for our grid, which considerably speeds up the computation.
        }
        \label{2D_equator_pole}
\end{figure}

However, as illustrated in \fig{2D_equator_pole}, for a grid with a finite size, the orientation of the grid can slightly affect the way the temperature is represented in the model, and consequently, the resulting spectrum.
We find that this effect is about 1~ppm.
For the rest of this section, we choose to place the star at the pole ($\latstar$~=~90\textdegree) because it allows us to use only one longitude ($\nlon = 1$) and one azimuthal angle ($\ntheta = 1$,
see \sect{3D_case}), which considerably speeds up computations.

\begin{figure}\centering
        \includegraphics[width=.5\textwidth]{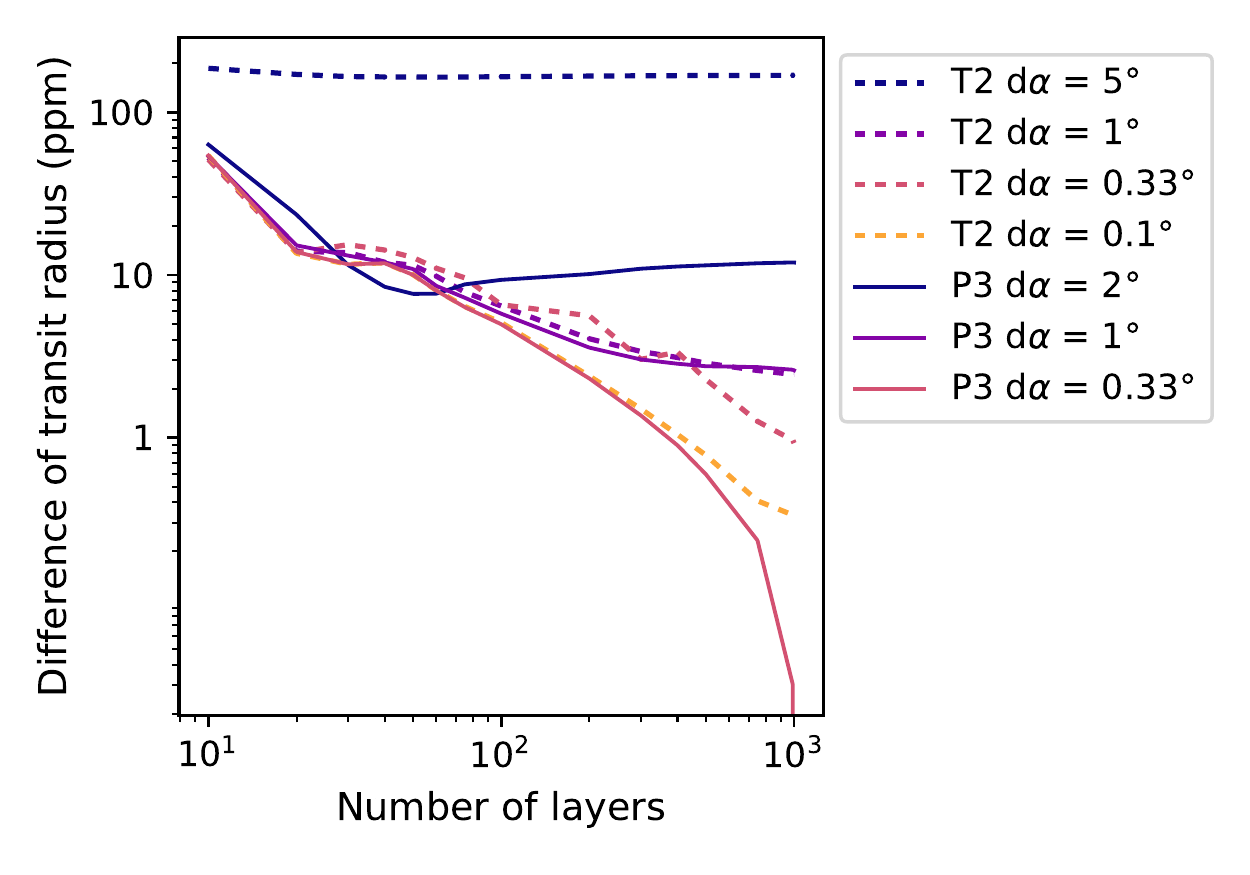}
        \caption{Convergence of \taurex 2D and \pytmo when the atmospheric grid resolution is increased.
        The Y-axis indicates the average of the absolute difference over all wavelengths considered between each model and \pytmo ($\nlayers, \nlat$) = (1000, 540).
        The X-axis indicates the number of layers $\nlayers$ in the model.
        The legend indicates the angular resolution of the model, $\diff\alpha$, equal to $\beta/\nslices$ for \taurex (T2) and $180/\nlat$ for \pytmo (P3).
        For example, $\diff\alpha = 1\degree$ leads to $\nslices = 10$ and $\nlat = 180$.
        \pytmo was run with $\latstar$ = 90\textdegree
        (see \fig{2D_equator_pole}),
        so that only one longitude and one azimuthal angle are necessary, i.e., $\nlon = \ntheta = 1$.
        }
        \label{taurex_convergence_2d}
\end{figure}

We can now compare the convergence of \taurex 2D with that of \pytmo.
The results are shown in \fig{taurex_convergence_2d}.
This figure shows the convergence of \taurex with an increasing number of slices and \pytmo with a increasing number of latitudes for an increasing number of layers.
With an equivalent angular resolution (e.g., $\nslices = 10$ and $\nlat = 180$, which leads to an angular width of $1$\textdegree~for each angular point), \taurex and \pytmo follow a very similar trend.
The models converge to a difference smaller than 1~ppm with a sufficiently high resolution.

\subsection{3D simulations}
\label{tests3D}

\begin{figure}\centering
    \includegraphics[width=.49\textwidth,keepaspectratio]{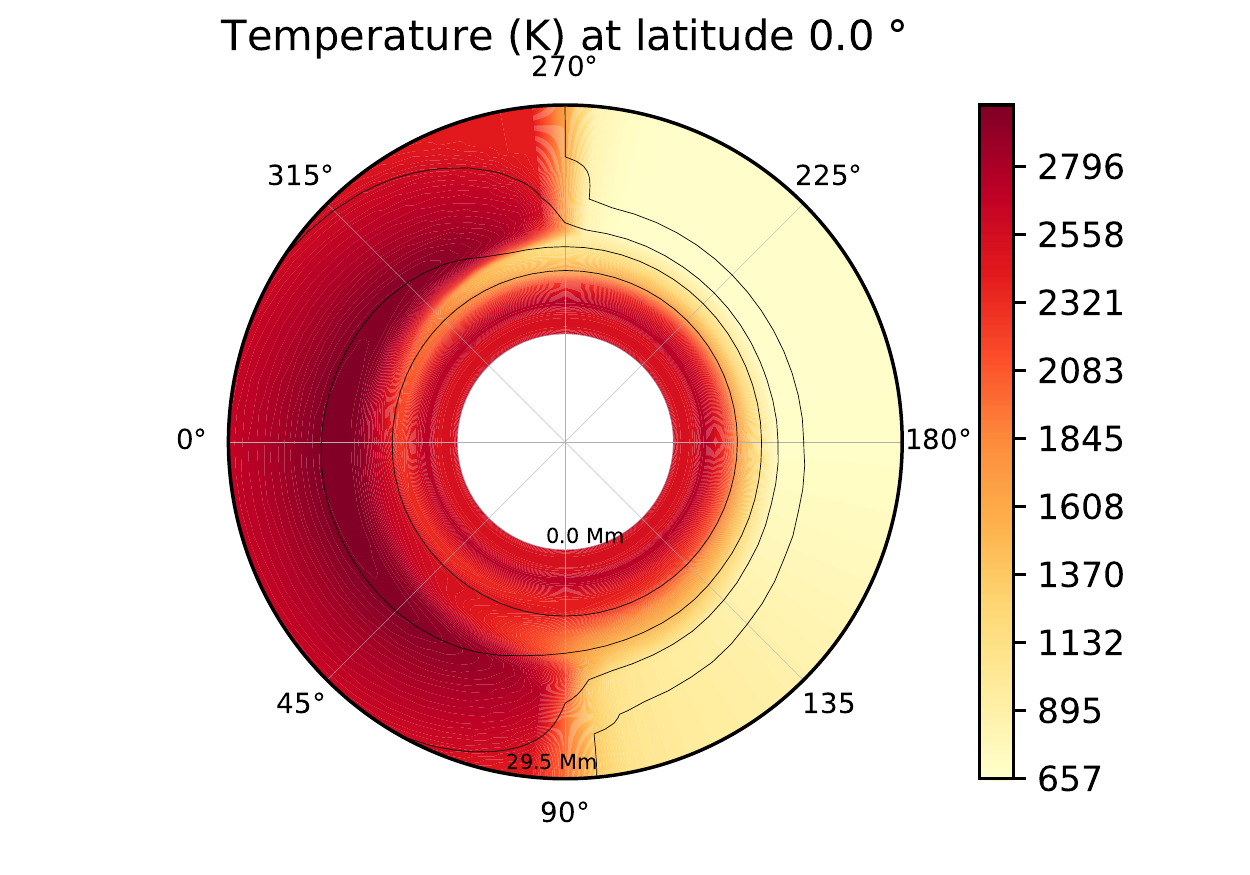}
        \label{fig:wasp121_equator}

        \includegraphics[width=.5\textwidth]{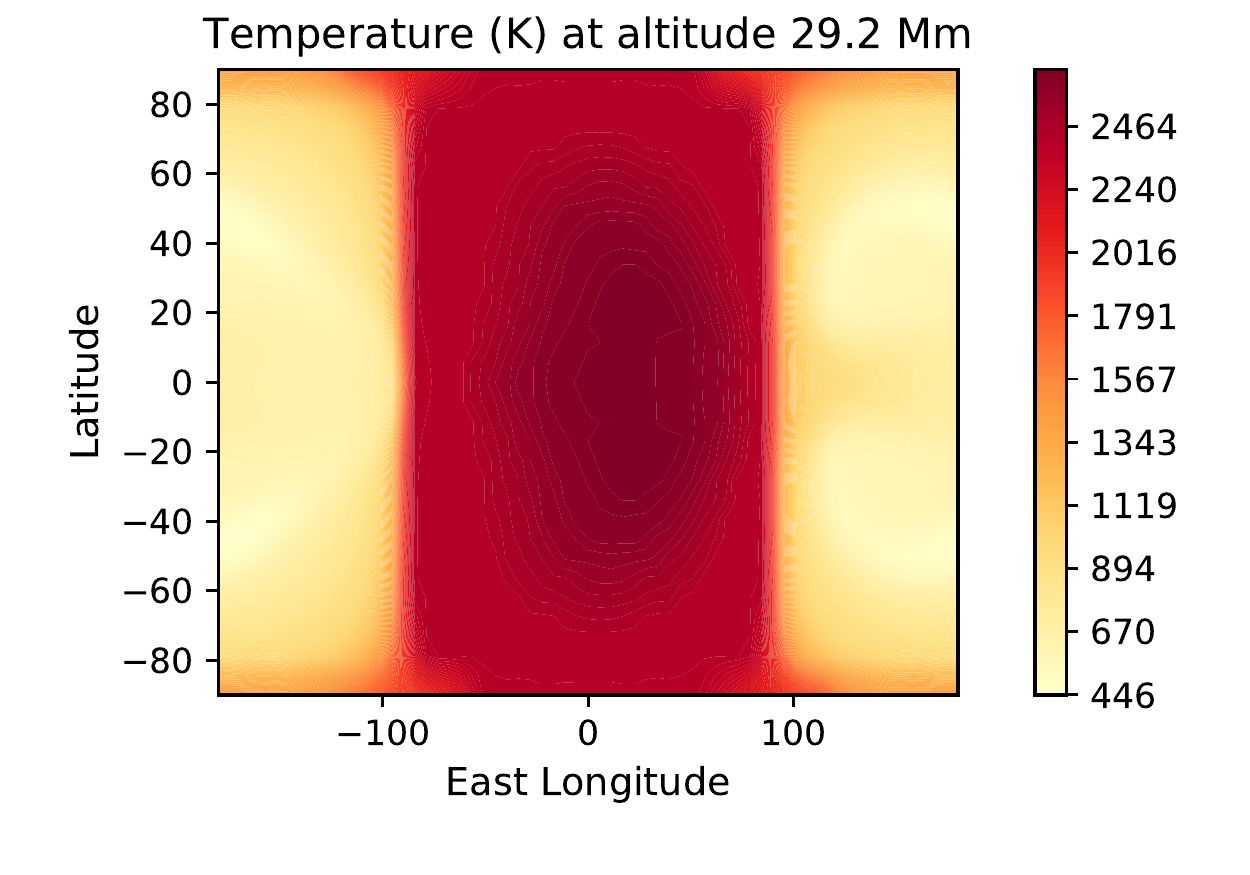}  \label{fig:wasp121_altitude}
        \caption{Temperature maps of \wasp.
        The top map shows an equatorial slice. The planet radius is divided by 10 for visual reasons. The black lines correspond to different isobar levels.
        The bottom map shows a slice at high altitude (29.2~Mm, i.e., 29200~km). The hottest point is slightly shifted to the east limb, i.e., the trailing limb.
        }
        \label{fig:wasp}
\end{figure}

We study here a 3D GCM simulation based on \wasp \citep{parmentier2018,pluriel2020strong}. One of the characteristics of this simulation is
the strong dichotomy between the temperature on the dayside and that of the nightside.
The simulation is shown in \fig{WASP} from the east
in the equatorial plane.
Temperature maps
in the equatorial plane
and at high altitude are given in \fig{fig:wasp}.
This simulation also has a slight east-west asymmetry; the hottest point is shifted toward the east. This feature is most visible in \fig{fig:wasp121_altitude}.
The chemistry is given by tables from \citet{parmentier2018} and includes He, H$_2$, H, H$_2$O, CO, TiO, and VO.

\begin{figure}\centering
        \includegraphics[width=.5\textwidth]{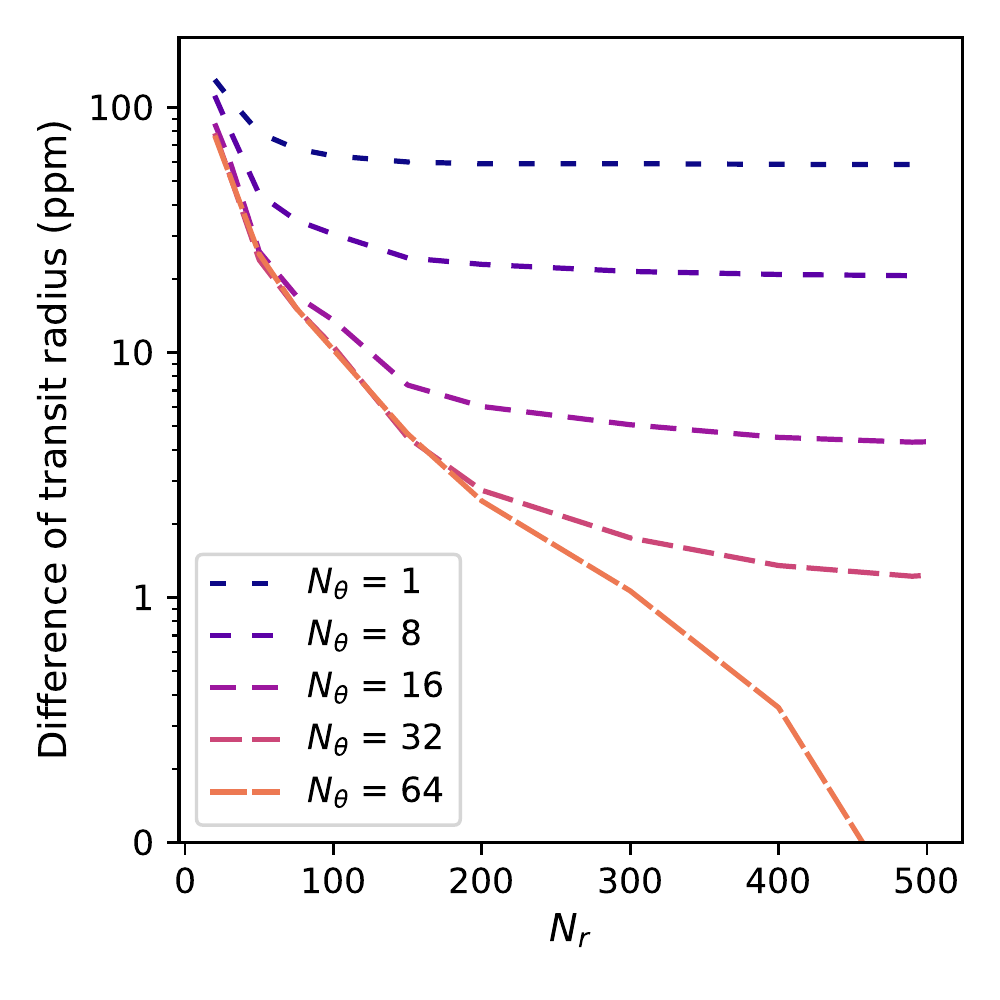}
        \caption{Convergence of \pytmo when the number of rays [$\nradial \times \ntheta$] crossing the atmosphere is increased,
        using as a reference point the converged model ([$\nradial \times \ntheta$] = [$500 \times 64$]).
        The simulation is a GCM with an equilibrium temperature $T_{eq} = 2100 K$ based on the characteristics of \wasp, and including TiO and VO.
        The number of cells in this simulation is $(\nlayers, \nlat, \nlon) = (100, 32, 64)$.
        }
        \label{pytmo_convergence}
\end{figure}

Using this (heterogeneous) 3D GCM simulation, the number of latitudes $\nlat$ and longitudes $\nlon$ is fixed.
The number of layers $\nlayers$ of the altitude grid may be chosen as different from that of the input simulation.
To simplify the number of parameters, we simply set this number to the number of radial points $\nradial$ in the polar grid of rays ($\nradial \times \ntheta$).
To study the accuracy of the model, we can therefore change the number of rays, as we show in \fig{pytmo_convergence}.
The model tends to converge if we have enough radial and angular points to correspond to the GCM resolution, that is, at least 32~angular points, although we might apparently need to double the number of radial points to obtain better accuracy.

\section{Performance and optimizations}
\label{model-time}

We now study the performance of our models and ways to reduce the time and memory required for the computation of a transmission spectrum.

\subsection{\taurex}

\taurex is mainly used for its retrieval feature.
We must therefore ensure that the forward model can be computed fast enough for the retrieval to be achieved in a reasonable amount of time for the user.

In its 1D variant, an important part of the arithmetic complexity of \taurex is the computation of the optical depth, which is related to the number of cells that the rays are passing through, and for which we must compute the formula of Eq.~\ref{eq:tau}.
The number of calculations in 1D leads to a complexity for the optical depth of
\begin{equation}
    \OC{\frac{\nlayers^2}{2} \cdot \nwl},
    \label{1D_absorption_complexity}
\end{equation}
where $\nwl$ is the number of wavelengths.
As a result of the assumption that day and night are identical, a ray at the altitude of level $i$ only intercepts the $\nlayers-i$ levels above $i$.
With $\nlayers$ rays, the number of layers for which the optical depth has to be computed is thus $\sum_{i = 1}^{\nlayers} i = \frac{\nlayers(\nlayers+1)}{2}$.
We focus here on molecular absorption and leave Rayleigh scattering and the continuum absorption out of this study.

In the 2D version of the model, however, we must account for the day- and nightside
(therefore doubling the number of computations of \eq{1D_absorption_complexity}),
as well as all the intersections of the layers with the slices in the linear transition around the terminator.
This leads to a complexity for the 2D optical depth of
\begin{equation}
    \OC{\left(\nlayers^2 + \nlayers \cdot \nslices\right)\cdot \nwl}.
    \label{2D_absorption_complexity}
\end{equation}
We show in the results that the main calculations in 2D are done for the number of opacities $\chi_{m,\lambda}[\alpha, r]$ that are to be interpolated.
The arithmetic complexity of computing the opacities is
\begin{equation}
    \OC{C \cdot \nlayers \cdot \nslices \cdot \nwl  \cdot \nmol},
    \label{2D_opacities_complexity}
\end{equation}
where $C$ is the cost of computing one interpolation, and $\nmol$ is the number of molecules.
Incidentally, adding molecules will increase the cost of the interpolation of the opacities (Eq.~\ref{2D_opacities_complexity}), but not of the optical depth itself (Eq~\ref{2D_absorption_complexity}).
Eq.~\ref{2D_opacities_complexity} is also valid in 1D, but in this case, $\nslices = 1$.

To ensure that the model behaves as expected, we performed a series of measures following the method we used and the dimensions of the atmospheric grid that were considered.
These measures are gathered in Tab. \ref{time_taurex}.
\begin{table}[h]\centering
      \begin{filecontents*}{img/taurex_timing.dat}
software,n_iter,n_layers,n_slices,beta,day_temp,night_temp,deep_temp,time_1D,t2_2,t2_10,t2_20,t2_30
TauREx,100,100,0,10,2500,3300,500,0.2593341986338297,0.5194946765899658,1.1987022479375204,1.9892582496007283,2.7997163693110148
TauREx,100,200,0,10,2500,3300,500,0.7786335229873658,1.7259700218836467,3.049527565638224,4.721514081954956,6.324856877326965
\end{filecontents*}
\pgfplotstabletypeset[
              every head row/.style={before row={%
              \hline & & \multicolumn{4}{c|}{$\nslices$ (2D)} \\
              },
              after row=\hline,
              },
    columns={n_layers,t1,2,10,20,30},
    columns/n_iter/.style={column name=$n_{iter}$,column type=|c},
    columns/n_layers/.style={column name=$\nlayers$,column type=|c},
    columns/n_slices/.style={column name=$\nslices$},
    columns/time_1D/.style={column name=1D},
    create on use/t1/.style={create col/expr={\thisrow{time_1D}}},
    create on use/2/.style={create col/expr={\thisrow{t2_2}}},
    create on use/10/.style={create col/expr={\thisrow{t2_10}}},
    create on use/20/.style={create col/expr={\thisrow{t2_20}}},
    create on use/30/.style={create col/expr={\thisrow{t2_30}}},
    columns/30/.style={column type=c|},
    columns/t1/.style={column name=1D,column type=c|,zerofill,precision=2,},
    multistyles={1,...,5}{,zerofill,precision=2,}
      ]{img/taurex_timing.dat}
    \caption{Average time (s) to run one \taurex model,
    considering the molecular absorption only (no Rayleigh scattering or CIA)
    of four molecules on 39124 wavelengths from 0.3 to 15~$\mu$m.
    }
    \label{time_taurex}
\end{table}
Because the whole calculation scales linearly with the number of wavelength points ($\nwl$) at which the model is run, it is set to a constant value (39124 wavelength points) for this study.

There are multiple interesting trends that we can observe in this table.
First, we can observe that there is a factor of 2 between the 1D and the 2D model with two slices because there is a day- and a nightside.
Second, the computational time seems to increase less than quadratically with respect to the number of layers in 1D, showing that there are enough molecules to make the interpolation of the opacities a significant part of the computations.
Third, for a large number of slices, the time also increases linearly with the number of slices, which means that the cost of one interpolation, $C$, in Eq.~\ref{2D_opacities_complexity} and the number of molecules $\nmol$ are large enough to make the interpolation of the opacities the dominant part of the computations.
Increasing the number of layers will change this behavior, as the quadratic complexity of Eq.~\ref{2D_absorption_complexity} will start to be dominant again.

A good compromise between accuracy and computational time seems to be running 2D models with 200~layers and 30~slices.

\subsection{\pytmo}

In 3D, the great majority of the calculations lies in the interpolation of the opacities.
This interpolation scales with the number of cells for which we need to compute the opacity.
To decrease the number of calculations, we can therefore first identify how the number of cells for which we need to compute an interpolation can be reduced.
The first step is to realize that we do not need (and cannot afford in terms of time and memory) to compute the opacity of all $\nlayers \times \nlat \times \nlon$ cells in the model. A large part of the cells is not crossed by any light ray, and therefore we do not need to compute their opacity.

A naive algorithm would simply iterate over all the rays and compute the opacity for each segment that is crossed by a ray.
However, depending on the resolution of the model, many cells may be crossed by multiple rays, so that we can reuse the information from one ray to another.
In addition, when we have the opacities of all segments of one ray, the optical depth of that ray can be computed, and the opacities may be discarded.
In light of these two facts, we developed an algorithm that will group light rays together at each azimuthal angle $\theta$.
This means we can benefit from the reusability of opacities within that angle (along the radial axis) while being able to release the memory of all opacities when the optical depth of all rays within this angle have been computed.
The memory will therefore almost be a constant throughout the execution of the program, if we consider each angle to be equivalent (this depends on the resolution of the grid).
This method, which we refer to as "Per-angle" in the experiments, works quite well for every kind of problem (completely heterogeneous to isothermal).

However, the computational cost can be decreased further by making a few assumptions.
Some atmospheric simulations may have a number of cells that have identical physical properties, for example, in the case of the 2D representation of Eq. \ref{temperature2D}.
If this number is sufficiently large, we can also aggregate these cells together to further reduce  the number of opacities that are to be computed.
This method is referred to as "Identical" in the experiments.

\begin{figure}\centering
                \includegraphics[width=0.48\textwidth]{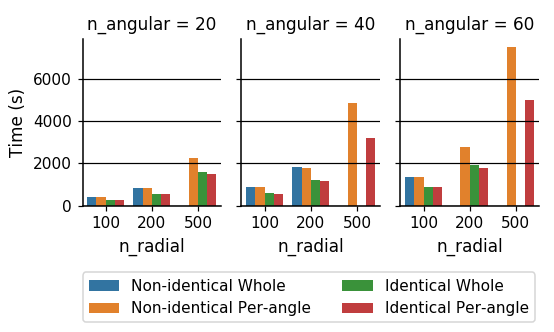}
                \caption{Time (s) required by \pytmo to compute the transmission spectrum of a 2D atmospheric grid defined with Eq. \ref{temperature2D}.
                Two methods are turned on or off: the Per-angle method (off is "Whole"), and the Identical method (off is "Non-identical").
                The number of rays $\nradial \times \ntheta$ = n\_radial$ \times $n\_angular also varies.
                Missing points have run out of memory.}
                \label{per_angle_times}
\end{figure}

\begin{figure}\centering
                \includegraphics[width=0.48\textwidth]{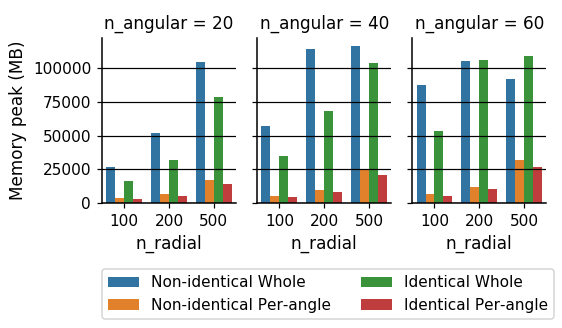}
                \caption{Real memory peak (MB) of the corresponding points in \fig{per_angle_times}. The machine has 128GB of RAM, and the recorded peak is below that point for those that have run out of memory (missing points in \fig{per_angle_times}).}
                \label{per_angle_memory}
\end{figure}

We compare these two methods in Figs. \ref{per_angle_times} and \ref{per_angle_memory} for a 2D problem defined with Eq. \ref{temperature2D}.
These two figures show the computational time required for a model to run and its memory consumption peak, respectively.

The first observation we can make here is the effect of the Identical method, which reduces the time and memory complexity of the program by approximately one-third in this case.
However, this is possible only due to the characteristics of the simulation, in which many cells are actually identical.
We must emphasize that a completely heterogeneous simulation would not benefit at all from this method.

A second observation we can make is the drastic memory saving due to the Per-angle method.
As we mentioned earlier, this method allows us to discard the opacities after each angle and keep the memory due to the opacities below a constant.
However, other variables such as the transmittance (if needed) will still increase in size with respect to the number of angles.

In conclusion, the Per-angle method provides a drastic memory reduction, while the Identical should be used only for data that contain redundancies.
A method to ensure that the computation will not run out of memory is also under study, as well as a parallel version.

\section{Examples of applications}
\label{sec:application}

\pytmo allows us to do more than just compute a more realistic spectrum from a static atmospheric structure.
The transmission signal from a planet is expected to vary over time due to two main reasons: (1) because the atmospheric structure itself is variable, whether it is the temperature, the composition, or the distribution of clouds and hazes;
(2) because the planet rotates during a transit (even when in synchronous rotation), showing us a slightly different cross section of its atmosphere.
In this section, we illustrate these two effects.

\subsection{Atmospheric variability}

\begin{figure}[ht]\centering
        \includegraphics[width=.49\textwidth,height=.7\textheight,keepaspectratio]{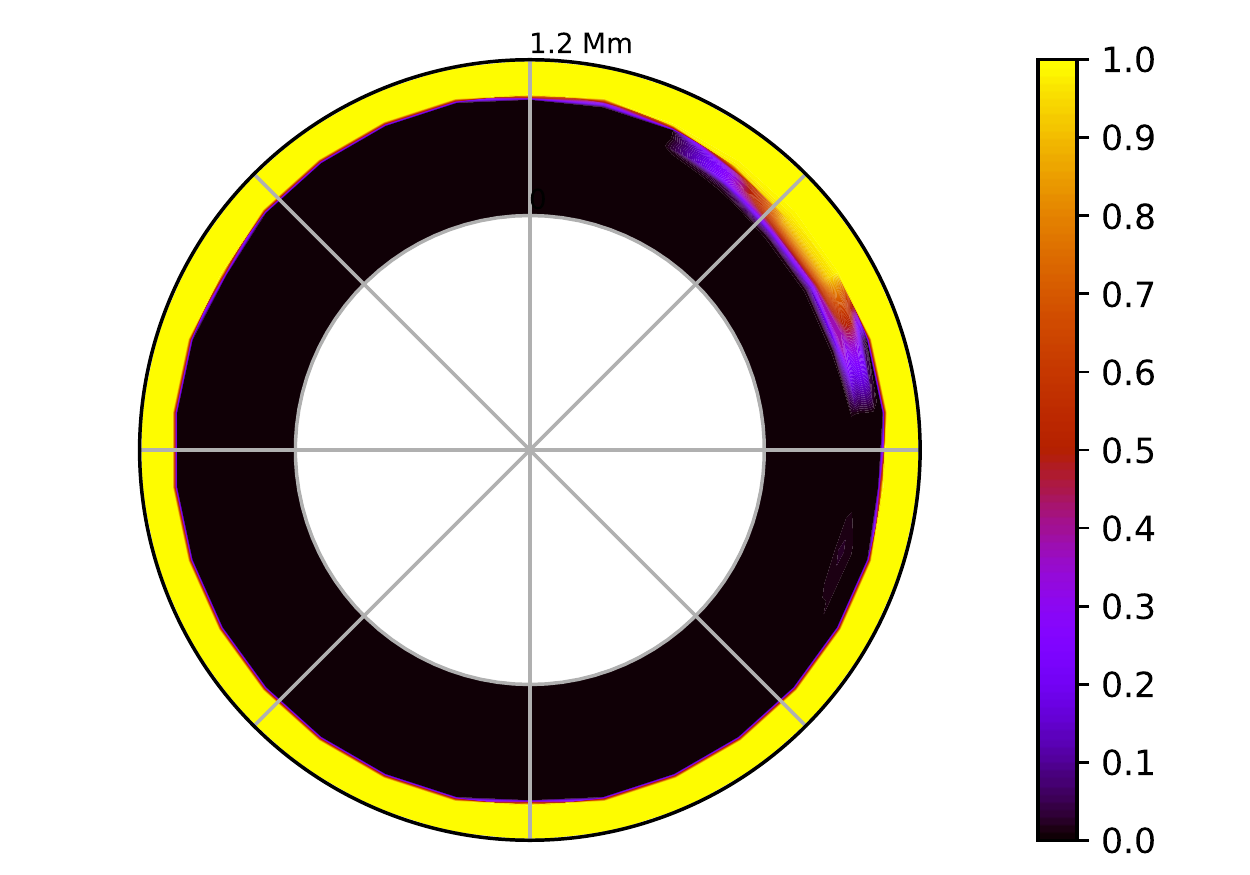}%

        \includegraphics[width=.49\textwidth,height=.7\textheight,keepaspectratio]{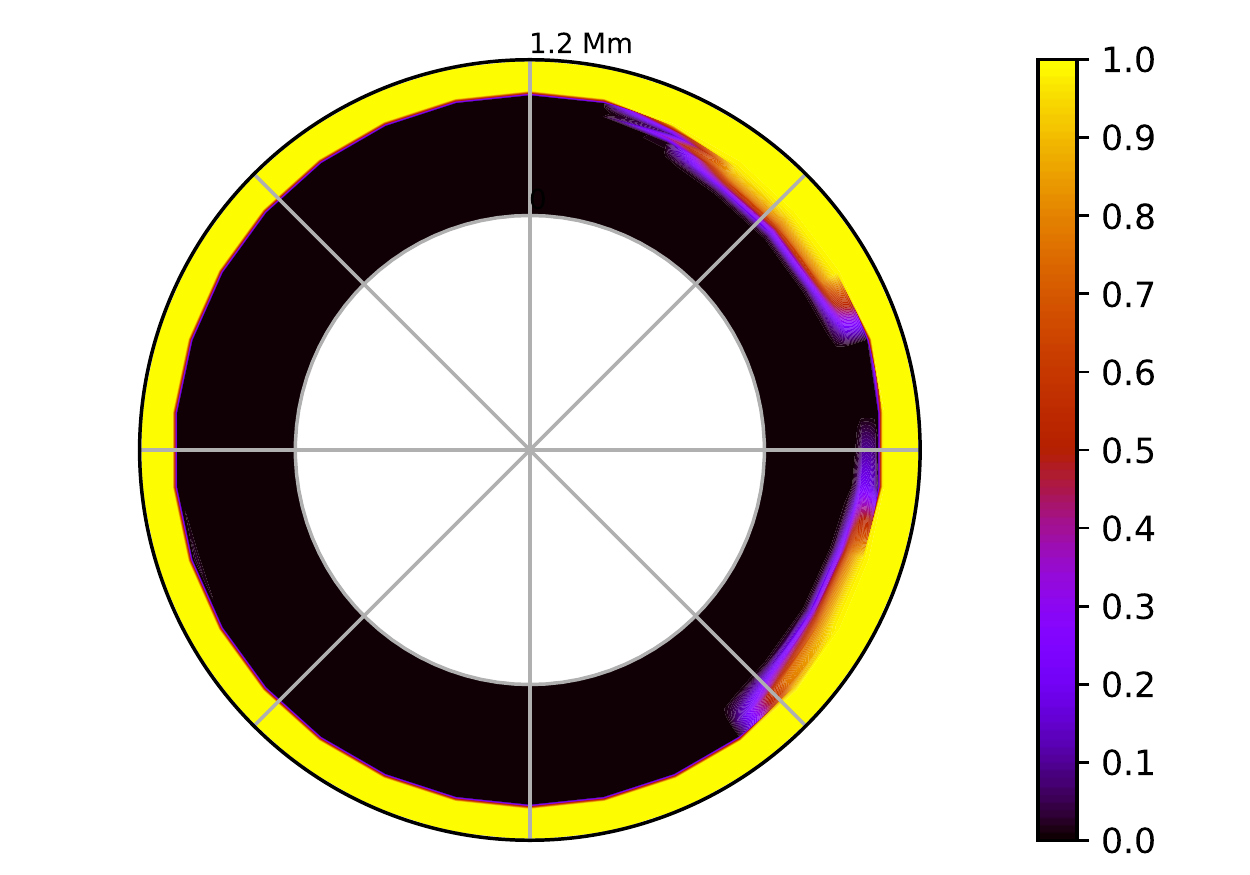}%
        \caption{Example of transmittance maps
        at two different times of a simulation
        of K2-18b \citep{charnay2020formation}
        at a wavelength of 0.6 $\mu m$ from the observer's point of view.
        The atmosphere is transparent when the transmittance is equal to 1, and  it is opaque when it is equal to 0. The absence of clouds is visible on the right side, which corresponds to the west limb.
        The atmosphere scale height has been multiplied by 10 for visual reasons.
        The altitude is in Mm, i.e., thousands of km.
        }
        \label{transmittance}
\end{figure}

In a recent study, \citet{charnay2020formation} investigated the dynamics of clouds on temperate mini-Neptunes, taking the example of K2-18b. They revealed that the abundance of clouds at the terminator was highly variable. To assess the effect of these clouds on the transmission spectrum, we ran \pytmo on the climate model results at various time steps. \fig{transmittance} shows transmittance maps calculated using \eq{eq:transmittance}.
These maps provide more information than the sole transmission spectrum that can be obtained by their spatial integration (see \eq{eq:integral}). Here we can use these maps to infer the effective fraction of the limb covered by clouds, as well as the difference in the effective absorption altitude for a clear versus cloudy atmsophere.

As the two maps in \fig{transmittance} correspond to two time steps in the simulation, we can also follow the movement of aerosols and the variation of the cloud fraction over time.

As the climate model is expected to predict a realistic evolution of the clouds in time, this can allow us to quantify how the variability of the cloud structure during a single transit could affect the observed signal.
\subsection{Rotation of the planet during transit}
\label{ssec:rotation}

An interesting aspect of our 3D model is also that the geometry of the observation can be changed. When a planet passes in front of its star, it rotates slightly, showing the observer a phase that varies with time. This occurs even when the planet is in a tidally synchronized rotation. In this case, the star (and the terminator) remains fixed in the reference frame corotating with the planet, but the observer (and the limb it probes) is moving.

In the case of an asymmetric and heterogeneous 3D atmosphere,
a change in the planet’s phase angle implies that
the light rays will not probe the same areas of the atmosphere. This therefore completely changes the associated transmittance map and the resulting transmission spectrum. Interestingly, this should create asymmetries in the transit light-curve \citep{EJ21}.

While we are in the process of implementing a full-fledged light-curve generator, we wish to quantify in a simple way here how much the spectral transit depth might vary due to these effects.

\begin{figure*}\centering
        \includegraphics[width=\textwidth]{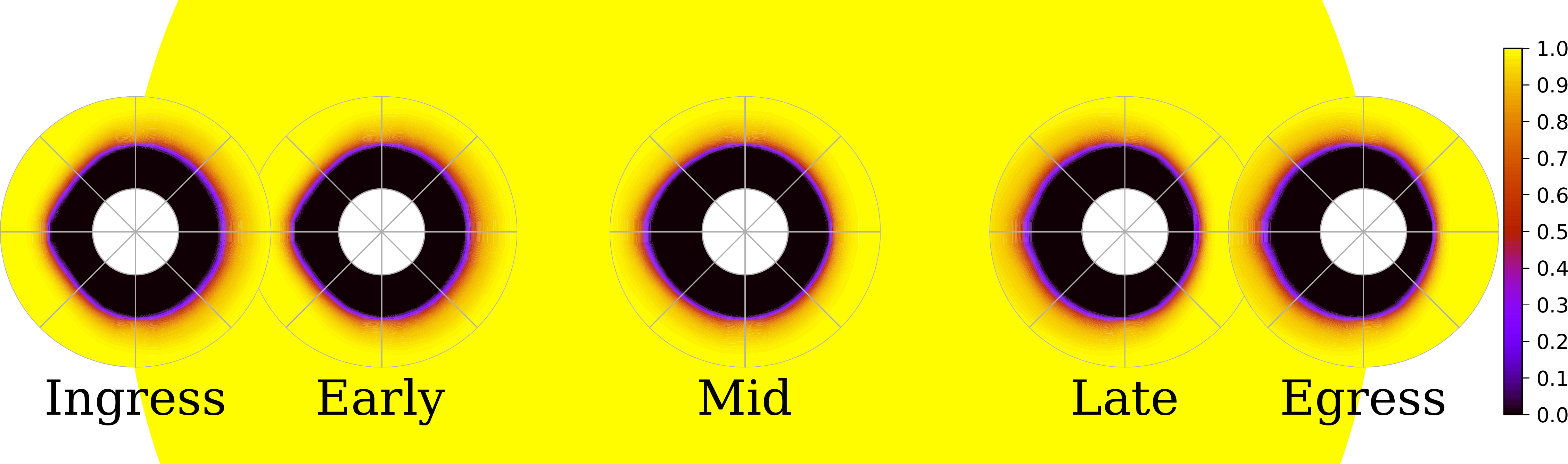}
        \caption{Transmittance maps of \wasp at 0.6 $\mu m$ for the five orbital phase angles whose spectra are shown in \fig{fig:wasp121_transit_effective_radii}.
        The orbital phase angle is $\phi = -15$\textdegree\xspace for ingress and $\phi = 15$\textdegree\xspace for egress.
        For visual reasons,
        the planet atmosphere has been enlarged with respect to its radius,
        and the early and late transmittance maps are slightly shifted.
        Only half the planet covers the star at ingress and egress.
        }
        \label{fig:wasp121_transit}
\end{figure*}

\begin{figure}\centering
        \includegraphics[width=.5\textwidth]{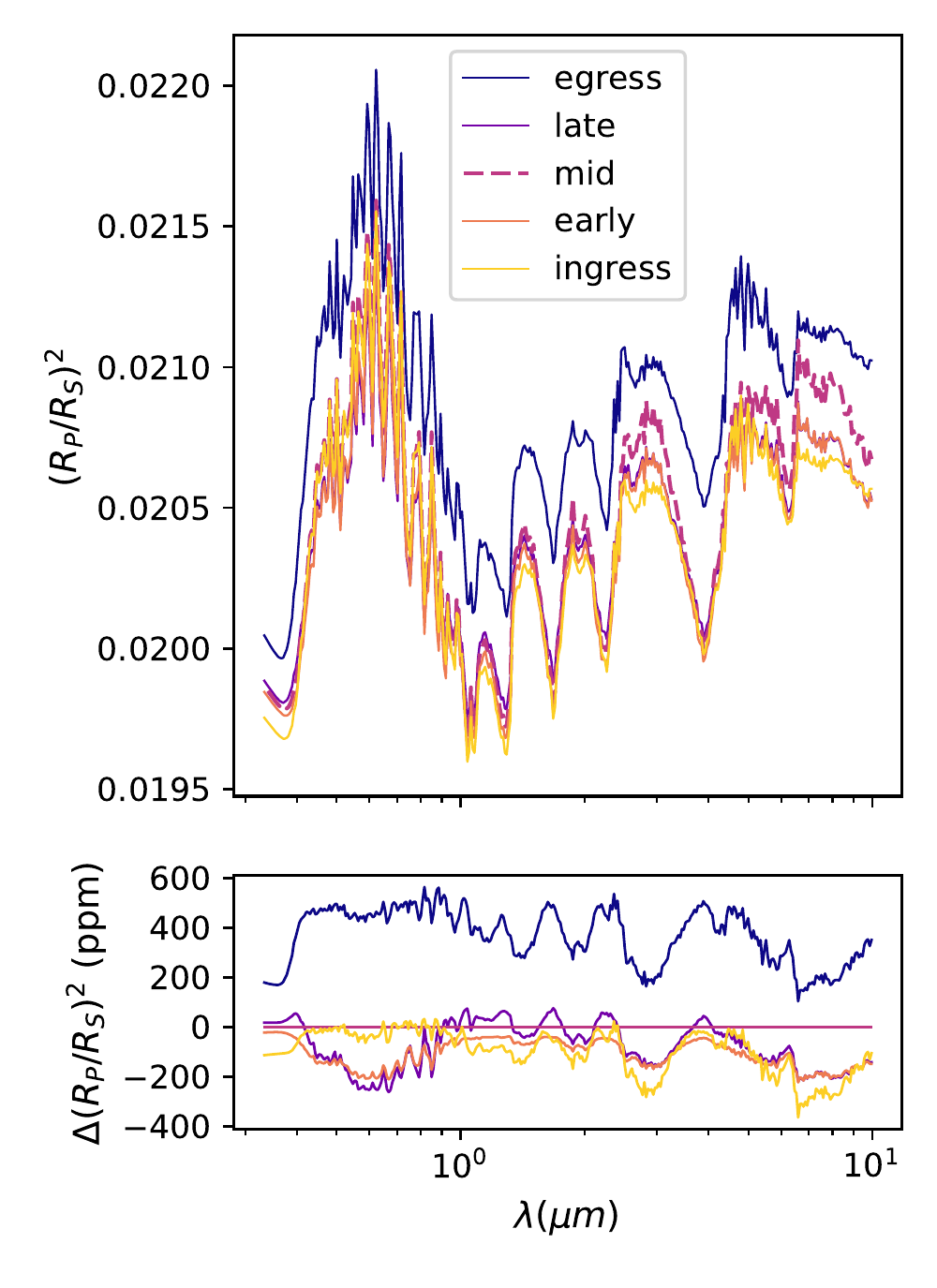}
        \caption{Spectral variations of the transit depth of \wasp during a transit for the phases listed in \fig{fig:wasp121_transit}.
        The bottom plot shows the difference between each spectrum and the mid-transit spectrum, taken as a reference.
        }
        \label{fig:wasp121_transit_effective_radii}
\end{figure}

To do this, we took the example of \wasp (\fig{fig:wasp}; details of the simulation can be found in \sect{tests3D}) and ran \pytmo at five different phases during the transit, as illustrated in \fig{fig:wasp121_transit}. The planet was considered to be tidally locked.
This figure shows the transmittance maps at different stages of the transit at a wavelength of $0.6~\mu$m.
It gives an example of how transmittance maps may evolve during a transit, and which information we can retrieve from them.
We only took the effect of the (synchronous) rotation of the planet between ingress and egress into account because we assumed a stationary atmosphere during the transit.
The selected phases are listed below.
\begin{enumerate}
    \item Ingress:
    Half of the planet (west limb) is in front of the star.
    For the system parameters we used, this corresponds to an orbital phase angle of $\phi = -15$\textdegree.
    The center of the planet is located at the edge of the star.
    \item Early: The planet has completed entering the transit (second contact; $\phi = -13$\textdegree).
    \item Mid: The planet is at mid-transit, $\phi = 0$\textdegree.
    \item Late: The planet is preparing to exit the transit, $\phi = 13$\textdegree.    \item Egress: The east limb is in front of the star. The planet is exiting the transit, $\phi = 15$\textdegree.
\end{enumerate}
\fig{fig:wasp121_transit_effective_radii} shows the relative transit depth (\eq{eq:integral}) for each phase. To facilitate comparison, we removed the effect of stellar limb darkening, even though it will be accounted for when computing realistic light-curves. For the ingress and egress spectra, only one limb is in front of the star, so that we multiplied the covered area by two to facilitate comparison.

The figure shows that these transit depth variations are not negligible: as an example, the early spectrum is 110~ppm below mid-transit on average (with a minimum at $-200$~ppm in certain wavelength regions, especially in TiO/VO bands,
i.e., from around 0.4 to $1~\mu m$).
The early and the late spectra follow similar patterns, including in the TiO/VO bands and higher wavelengths, indicating a general symmetry in the TiO/VO repartition.
In our simulation, TiO and VO are mainly located around the terminator.
The decrease in absorption in early and late spectra (with respect to mid-transit) is due to the rotation of the terminator disk, which shows the highest cross section at mid-transit.

As shown in
\fig{fig:wasp121_altitude}, the temperature structure is not completely symmetrical because there is an eastward shift of the hottest region of about $20^{\circ}$, even though the day-night transition remains very sharp and symmetric. This eastward shift of the hot spot results in an east limb that is larger than the west one, which is visible in all transmittance maps.
The elongation is less visible during ingress as the hot spot is behind the planet from the observer, and this is exacerbated during egress as the spot is closer to the east limb.

During the early part of the transit, the limb is different from the terminator (rotated by $-\phi$), and we probe deeper in the dayside west of the planet and deeper in the nightside east of the planet.
The situation is exactly reversed at the late position.
\fig{fig:wasp121_transit} shows that the transmittance map at the early position is mostly symmetrical because the asymmetry of the temperature map is compensated for by the rotation of the planet during its orbit.
Then, moving to the late step, the reverse situation occurs: The temperature on the east limb is hotter, implying a greater scale height, but the west limb is colder, inducing a smaller scale height.
The temperature eastward shift seems to lead to a difference smaller than 100~ppm (the largest spectral difference).
The location of the molecules (around the terminator) decreases this difference further, and the two spectra are very similar in most parts (with an average
difference
of less than 40~ppm).
Although the changes from the mid-transit to the early and the late transmittance maps are (longitudinally) reversed, these differences disappear when the transmittance maps are spatially integrated to generate the spectra (\eq{eq:integral}).

For ingress and egress, only the planet and atmospheric half that is in front of the star (the west and east limbs, respectively) are considered for the integration of the transmittance into a spectrum.
The rotation of the planet at egress means that the east limb of the planet is hotter (accentuated by the eastward shift of the temperature map), while the west limb is colder.
Because the east side alone is considered, the scale height of the atmosphere is larger and we observe a larger effective radius (see \fig{fig:wasp121_transit_effective_radii}).
This leads to stronger spectral differences with an average of 370~ppm and peaks up to 560~ppm.

For the ingress, the eastward shift of the hottest region (see \fig{fig:wasp121_altitude}) and the orbital phase of the planet implies that the light crosses colder regions of the atmosphere on the west limb.
This results in a smaller effective radius because of a smaller scale height.

The key points of this study are therefore threefold:
\begin{enumerate}[topsep=0pt]
    \item The rotation of the planet during transit results in variations of the transit depth of up to 300\,ppm for a hot Jupiter such as \wasp.
    \footnote{The egress and ingress spectra were multiplied by 2 to compare them to the other phases, therefore the differences shown in \fig{fig:wasp121_transit_effective_radii} for these phases are twice larger than for the real signal.}
    It should therefore be detectable.
    The noise in observations from the HST currently reaches around 50~ppm and can be as low as 20~ppm at best. This noise could be lowered to 10~ppm or less with the upcoming JWST \citep{Greene_2016}.
    \item The most important differences are between ingress and egress (when only half the planet covers the star) and are mainly due to the asymmetry caused by the eastward shift of the hot spot.
    \item Measuring these light-curve asymmetries would allow us to place constraints on the rotation of the planet and/or the direction of the hotspot shift without the need for a complete and expansive phase-curve.
\end{enumerate}

\subsection{Toward a time-domain analysis}
\label{sec:discussion}

As \pytmo can simulate any position for the observer, it can provide spectra and transmittance maps at any position during a transit.
The transmittance maps can be used to extract the part of the atmosphere (and planet) that covers the star, for example, during ingress and egress (see Sec. \ref{ssec:rotation}).
This information can also be used in the future for a time-domain analysis of the transit with \pytmo.
We are in the process of extending \pytmo to generate transit light-curves, which could be very useful to theoretically study transit observations. As \pytmo is fully 3D, the light-curves that are generated would be the closest to a real observation and would avoid biases due to 1D model assumptions. It will be interesting to compare the information extracted from light-curves by other codes \citep{Kreidberg_2015,Tsiaras2018,EJ21,Feliz2021} to the input model provided to \pytmo.

\section{Conclusion}

We have discussed the computation of transmission spectra for exoplanetary atmospheric simulations with a varying number of dimensions.
This method, implemented in \pytmo, handles atmospheric simulations with up to three dimensions, including GCM models.
The 2D formulation has also been integrated into the (initially one-dimensional) \taurex framework \citep{Al-Refaie2019}.
We then discussed the computational requirements and efficiency of each model, which is especially critical in the context of retrievals, as well as possible applications.

We have introduced a new version of \href{http://perso.astrophy.u-bordeaux.fr/~jleconte/pytmosph3r-doc/index.html}{\pytmo} that is more robust and flexible, which is open-source and under a BSD license$^{\ref{pytmodoc}}$.
Taking into account the 3D structure of the atmosphere during a transit is essential for generating consistent observations because assuming atmospheres to be homogeneous leads to strong differences in the transmittance maps and in the final integrated spectrum \citep{caldas2019effects, pluriel2020strong}.
We will further study the biases due to the 1D assumption of retrievals with \pytmo~2.0 in the second part of this series of articles, and quantify these biases.

The two-dimensional model was shown to be a good compromise between accuracy and computational requirements, making it a valid forward model for a retrieval.
Thanks to this method, we can remove the biases that were observed when a 1D forward model is used to retrieve very hot exoplanets.
We will discuss the relevance, precision, and reliability of this 2D retrieval in the third part of this series.

However, it should be noted that there might be other ways to parameterize 2D retrievals (discussed in Sec. \ref{introduction}). For instance, we know that east-west effects might also bias transmission spectra in warm atmospheres \citep{MacDonald2020}, where the jet stream cools down the west limb and heats up the east limb. In these configurations, our 2D model, which is
symmetric with respect to the star-observer line,
would not be able to give a better solution than a 1D model because its limb is homogeneous by definition.
2D retrievals using adapted configurations depending on the type of observed exoatmospheres are required. We could also develop hybrid 2D models that would take several geometric effects into account, keeping in mind that too many parameters in a retrieval code may create degeneracies.
Overall, \taurex~2D can infer the atmospheric parameters of specific exoplanetary types, that is, ultra hot Jupiters, with a good compromise between computational time and model precision.
This 2D version of \taurex will be made publicly available in the near future.

\begin{acknowledgements}
We are grateful to all of the \taurex developping team. This project has received funding from the European Research Council (ERC) under the European Union's Horizon 2020 research and innovation programme (grant agreement n$^\circ$679030/WHIPLASH). We thank the Programme National de Planétologie (CNRS/INSU/PNP) and the CNES for their financial support.
\end{acknowledgements}

\bibliographystyle{aa}
\bibliography{biblio}

\end{document}